\documentclass[10pt, conference, compsocconf]{IEEEtran}
\IEEEoverridecommandlockouts
\usepackage{times}
\usepackage{epsfig}
\usepackage{graphicx}
\usepackage{amsmath}
\usepackage{amssymb}
\usepackage{url}
\usepackage{multirow}
\usepackage{authblk}
\usepackage{color}
\usepackage{hyperref}

\ifCLASSINFOpdf
\else
\fi
\begin{document}
\title{Characterizing Data Analysis Workloads in Data Centers}
\author[1,2]{Zhen Jia}
\author[1,2]{Lei Wang}
\author[1*]{Jianfeng Zhan \thanks{* The corresponding author is Jianfeng Zhan.}}
\author[1]{Lixin Zhang}
\author[1]{Chunjie Luo}
\affil[1]{State Key Laboratory Computer Architecture, Institute of Computing Technology, Chinese Academy of Sciences}
%\affil[1]{Institute of Computing Technology, Chinese Academy of Sciences}
\affil[2]{University of Chinese Academy of Sciences, China}
%\affil[]{jiazhen@ncic.ac.cn, wl@ncic.ac.cn, \{zhanjianfeng, zhanglixin\}@ict.ac.cn}
%\href{\{jiazhen, wl\}@ncic.ac.cn, \{zhanjianfeng, zhanglixin, luochunjie\}@ict.ac.cn}
\affil[]{\{jiazhen, wl\}@ncic.ac.cn, \{zhanjianfeng, zhanglixin, luochunjie\}@ict.ac.cn}

\maketitle

\begin{abstract}
As the amount of data explodes rapidly, more and more corporations are using data centers
to make effective decisions and gain a competitive edge.  Data analysis applications
play a significant role in data centers, and hence
it has became increasingly important %for system designers
to understand their
behaviors in order to further improve the performance of data center computer systems.
%However, to the best of our knowledge,
%there is no public benchmark suite that covers diversity of data center workloads and
%can be used to study data center
%applications.

In this paper, after investigating three most important application domains in terms of page views and daily visitors,
 %\emph{search engine, social networks, and electronic commerce},
 we choose eleven  representative data analysis workloads and characterize their micro-architectural characteristics   by using hardware performance counters,
in order to understand the impacts and implications of data analysis
workloads on the systems equipped with modern superscalar out-of-order processors.
%behind creating such a benchmark suite and then present the proposed benchmark
%suite, \emph{HVCBench}.  Our benchmark suite is extracted 21
%representative applications selected from massive data center applications.
Our study on
the workloads reveals
%two unique aspects of data analysis applications in data center.
%First,
that data analysis applications
%may have very different architectural characteristics, according which we give some
%implications to system engineers and processor researchers.
%they
share many inherent characteristics,
which place them in a different class from desktop (SPEC CPU2006), HPC (HPCC),  and service workloads, including traditional server workloads (SPECweb2005) and scale-out service workloads (four among six benchmarks in CloudSuite), and accordingly we give several recommendations for architecture and system optimizations.

On the basis of our workload characterization work, we released a benchmark suite named \emph{DCBench} for typical datacenter workloads, including data analysis and service workloads,  with an open-source license on our project home page on \url{http://prof.ict.ac.cn/DCBench}. We hope that \emph{DCBench} is helpful for performing  architecture and small-to-medium scale system researches for datacenter computing.
\end{abstract}


% \fi
\IEEEpeerreviewmaketitle

\section{Introduction} \label{intro}

In the context of digitalized information explosion,
%The amount of digitalized information is exploding rapidly.
more and more
businesses are analyzing massive amount of data -- so-called big data --
with the goal of converting big data to ``big value''. Data center workloads can be classified into two categories: services  and data analysis workloads as mentioned in \cite{zhan2012high} and \cite{barroso2009datacenter}.
%Many organizations have their own data centers in order to store and
%process big data. %either to use it internally, referred to as private clouds \cite{armbrust2009above}, or lease it to external customers, referred to as public clouds \cite{armbrust2009above}.
%Such as the Hadoop usage report \cite{HadoopUsageReport} lists over 160 companies or organizations using Hadoop clusters including
%Amazon, eBay, Yahoo etc. %whose scale range from several nodes to thousands of nodes.
%systems with up to hundreds of nodes and hundreds of petabyte of
%data. Considering that there are many other platforms besides MapReduce, such as
%Dryad by Microsoft, the actual number of organizations with large data centers
%is believed to be much higher than 150.
%As data volume increases, data analysis workloads can account for a large
%proportion in data centers workloads.
Typical data analysis workloads include business intelligence, machine learning, bio-informatics, and ad hoc
analysis~\cite{InfrastructureAtFacebook}\cite{HadoopUsageReport}.
%For example, Facebook analyzes more
%than 15 PB of data (2.5 PB after compression), including over 60 TB of new data
%(10 TB after compression) every day~\cite{InfrastructureAtFacebook}, and its applications range from plain event
%collection tools like insights for the advertisers, to more sophisticated
%utilities like friend recommendation for normal
%users.
%For example, Ebay has more than 97 million world-wide active
%buyers and sellers and 200 million items in more than 50,000 categories. It
%handles about two billion page views and 250 million queries every day \cite{ebayhic}.
%It has always been a dream of the sellers to recommend the right products to the
%right buyers %by mining the history data.
%The recommendation system, which a typical data analysis
%application with huge financial implications, aims at recommending the right products to the
%right buyers by mining user behavior and other logs.

The business potential of the
data analysis applications is a driving force behind the design of
innovative data center systems including both hardware and software\cite{zhancost} \cite{wang2012cloud} \cite{sang2012precise}.
For example, the recommendation
system is a typical example with
huge financial implications, aiming at recommending
the right products to the right buyers by mining user behavior
and other logs.
For  data analysis in data centers is a relatively new but very important application area, there is a need
to understand  various representative workloads' performance characteristics
and what optimizations will further improve the performance.
So characterizing data analysis
workloads  is meaningful for system designers and researchers to
gain insights on optimizing data center computer systems.

There have been much work proposed to evaluate data mining algorithms
or so-called scale-out cloud workloads in different aspects, such as MineBench \cite{narayanan2006minebench},
HiBench \cite{huang2010hibench},
and CloudSuite \cite{ferdman2011clearing}.
The state-of-the-art work of characterizing data center workloads on a micro architecture level is CloudSuite \cite{ferdman2011clearing}.
However, CloudSuite is biased towards online service workloads: among six benchmarks, there are four scale-out service workloads, (including \emph{Data Serving, Media Streaming, Web Search,  Web Serving}), and only one data analysis workload---{\em Naive Bayes}.  Our work
show the data analysis workloads are significantly diverse in terms of both speedup performance (Section \ref{why_not}) and micro-architectural characteristics (Section \ref{characterization_BDbenchmarks}).
%, implied that these applications have different
%user observation characteristics. In our experiments, some applications' speed up at 8 nodes are greater than 8, this is
%because that we use the same data set for all of the experiments, which resulted in the data set(around 150GB ) maybe too large for
%one slaves to process.
 %Meanwhile our  experiments in Section \ref{characterization_BDbenchmarks} also shows that different data analysis applications have
%different .
In a word, only one application is not enough to represent various categories of data analysis  workloads.

\begin{figure}
\centering
\includegraphics[scale=0.7]{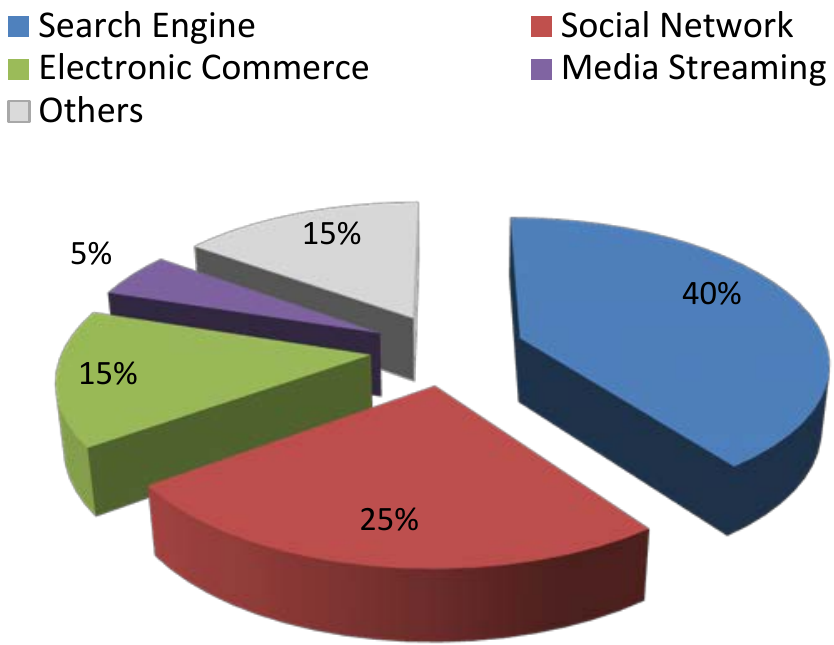}
\caption{Top sites in the web \cite{Alexa}.}\label{share}
\end{figure}

In this paper, firstly we single out three important applications domains in Internet services:  \emph{search engine, social networks, and electronic commerce}
(listed in Figure \ref{share}) according to
a widely acceptable metrics --- the number of page views and daily visitors. And then, we choose eleven representative
data analysis workloads (especially intersection workloads) among the three applications domains.
%For those applications, we
%use a much large size input data set (greater than total working nodes' memory size) to drive them.
In comparison with that of CloudSuite described in \cite{{ferdman2011clearing}},
our experiment approach is more pragmatic.
First, we deploy a larger
input data set varying from 147 to 187 GB that are stored in both the memory and disk systems
instead of completely storing data (only 4.5GB for \emph{Naive Bayes} in \cite{ferdman2011clearing}) in the memory system. Second,
for each workload, we collect the performance data of
the whole run time after the warm-up instead of a short period (180 seconds in \cite{ferdman2011clearing}).
%We perform
%an extensive study of these applications.

Our study reveals that data analysis applications share many inherent characteristics, which place them in a different class from desktop (SPEC CPU2006), HPC (HPCC), and traditional server workload (SPECweb2005), and scale-out service workloads (four among six benchmarks in ClousSuite).
Meanwhile, we also observe that the scale-out service workloads (four among six benchmarks in CloudSuite) share many similarity in terms of micro-architectural characteristics  with that of  SPECweb2005,
so in the rest of this paper, we just use the service workloads to describe them. On the basis of our workload characterization work, we released
  a benchmark suite named \emph{DCBench} for datacenter computing.
%We will investigate more workloads to confirm this observation.
%and scale-out service
%have different micro architecture level behaviors with service workloads in data center.
%may have very different architectural characteristics, according which we give some
%implications to system engineers and processor researchers.
%they
%The data analysis workloads also share many inherent characteristics,
%In this paper, we do not distinguish between traditional service workloads and scale-out workloads and all call them service workloads for they are all different from data center analysis workloads in terms of micro architecture behaviors.
%share many inherent characteristics,
%which place them in a different class from desktop, HPC,  traditional server and scale-out service
%workloads, and accordingly we give several recommendations for architecture and system optimizations.
%workloads, characterized by CloudSuite.
Our key findings are as follows:
%\begin{itemize}

First, the data analysis workloads have higher IPC than that of the services workloads
%that are characterized by CloudSuite
%and traditional web server workloads, e.g., \emph{SPECweb},
 while lower than that of computation-intensive HPCC workloads, e.g., \emph{HPC-HPL, HPC-DGEMM}.

%Meanwhile data analysis workloads involve less intensive operating system (OS) calls than that of services workloads while more than that of HPCC workloads, SPECFP, and SPECINT workloads.

 Second, corroborating previous work \cite{ferdman2011clearing}, both the data analysis workloads and  service workloads suffer from notable front-end stalls, which may be caused by two factors: deep memory hierarchy with long latency  in modern processor \cite{ferdman2011clearing}, and  large binary size complicated by  high-level language and third-party libraries.
 However, we note the significant differences between the data analysis workloads and the service workloads in terms of stall breakdown:
the data analysis workloads suffer more stalls in the out-of-
order part of the pipeline (about 57\% on average), while the service workloads
suffers more stalls before instructions entering the out-of-order part (about 73\% on average).
 %This observation indicates the service workloads will not benefit from aggressive or even modest  degree of superscalar out-of-order execution.

 Third, %improving the L1 instruction cache and TLB hit ratios can promote the performance of the data analysis workloads, especially the service workloads.
%by reducing latency that pipeline back end waiting for instruction coming.
%So the L1 instruction cache and instruction TLB become focus.
the third-party libraries and high-level languages used by the data center workloads may aggravate the inefficiency of L1 instruction cache and instruction TLB. So when
writing the program (with the support of third-party libraries and high-level languages), the engineers should pay more attention to the code size.

 Fourth, for the data analysis workloads, L2 cache is acceptably effective, and they have lower
L2 cache misses (about 11 L2 cache misses per thousand instructions
on average) than that of the service workloads
(about 60 L2 cache misses per thousand instructions on average)
while higher than that of the HPCC workloads.
%It only	has more L2 cache misses than HPCC workloads.
Meanwhile, for the data analysis and service workloads, on the average 85.5\% and 94.9\% of L2 cache misses are hit in L3 cache, respectively. Considering  modern processors dedicate approximately half of the die
area to caches, optimizing the  LLC capacity  will improve the energy-efficiency of processor and save the die size.  For the service workloads, our observation corroborate the previous work \cite{ferdman2011clearing}: L2 cache is ineffective.
%reduced LLC capacity along with the removal of the ineffective L2 cache would offer
%access-latency benefits while at the same time freeing up die area
%and power .

%\emph{\textbf{Finally, for the data analysis workload, the misprediction ratio is lower than that of  most of the other workloads, so a simple branch predictor is preferred so as to save power and die area.}}

Finally, for the data analysis workloads, the misprediction
ratios are lower than that of most of the service workloads, which implies that the branch predictor
of modern processor is good. A simpler branch predictor may be preferred so as to save power and die area.

%The contributions of our paper are as follows.

%\begin{itemize}
%  \item Present our methodology for selecting representative applications in
%    different data analysis application domains.
%  %\item Propose a benchmark suite containing 21 benchmarks written in
%    %four different programming models.
%  \item Characterize the selected 11 workloads and compare them with
%    some famous benchmark suites.
%	\item Give some implications to system engineers and researchers according to our characterizations.
%  %\item Demonstrate that the diversity of data analysis applications
%   % necessitates the needs to include the 11 when do characterization %selected benchmarks.
%\end{itemize}

The remainder of the paper is organized as follows. Section \ref{requirements}
describes our benchmarking methodology and design decisions. %how we chosen representative workloads for data analysis applications.
%Section \ref{ben_over} describes the proposed benchmark suite.
Section \ref{character} states our experiment methodology.
Section \ref {characterization_BDbenchmarks} presents the micro-architectural characteristics of the data analysis workloads in comparison with other benchmarks suites.
% and gives some implications.
%Section \ref{implication}
Section \ref{dcbench} introduces our DCBench.
Section \ref{conclusion} draws conclusions of the full paper and mentions the future work.

%throughput oriented computing

%\section{Motivation}\label{section2}

%The goal of our work is to build a benchmark suite that characterizes data
%analysis applications in data centers.

%In this section we will first present the requirements for such a benchmark
%suite. Then, we discuss why the existing benchmark suites fail to capture these
%requirements.
\section{Benchmarking Methodology and Decisions} \label{requirements}
%\section{The Selected Data Analysis Applications in Data Centers} \label{requirements}

%A benchmark suite must have a target class of machines~\cite{bienia2008parsec}
%and a target class of applications.For this effort, it is data center systems
%. Such a benchmark suite must meet the following goals.
\subsection{Workloads Requirements} \label{reforus}
The workloads we choose must meet the following requirements.
%\begin{description}
%  \item[Being representative] is the first goal of a good benchmark suite.
%    Because data center applications can go from simple reporting to much more
%    sophisticated algorithms, a large pool of applications must be considered,
%    evaluated, and categorized before representative ones can be selected.
%    Because data center applications typically run with large scale systems,
%    being representative also means selecting input data that can not only
%    represent data in scale but also be deployed without first building a
%    data center.  Because workloads in data centers can change frequently
%    (workload churns~\cite{barroso2009datacenter}), being representative also
%    means including recently developed and emerging algorithms.
%
%  \item[Being comprehensive] is another one of the goals. While its main
%    meaning is to select at least one application from each categorized groups,
%    it also means to take consideration of all the programming models frequently
%    used by target applications. The four common programming models for big
%    data applications are MapReduce, Dryad, ,all-pairs and WorkQueue.  Our experiments in
%    Section \ref{character} will show that they are indeed different.
%\end{description}

  \textbf{Representative}
	
	There are many representative workloads in many specific fields, such as SPEC CPU for processors, SPECweb for Web servers. Those workloads are representative in their own fields. For data analysis field in data center,
  we should choose representative workloads in the most important application domains.
  %applications range from simple reporting to deep data mining, and only those workloads can reflect the real performance in data center.

	\textbf{Distributed} %The target class of machine is data center.
	
	%Data center is a large distributed environment, e.g., Google's data center has about 450000 nodes \cite{googlenode}; Facebook's Hadoop cluster has more than 2250 nodes \cite{Facebook_node}.
	In general, most of data analysis workloads in data center are distributed on several nodes for
	the large amount of data can not be processed efficiently in a single node.
	An application running on a single computer can not represent the applications in real world data centers.
   %be regarded a real world workload in terms of both programming model and input data set. So the workload in the benchmark suite should be distributed.

	\textbf{Employ State-of-art Techniques \cite{bienia2008parsec}}
	
	In data centers, workloads change frequently, which Barroso et al. call workload churns  \cite{barroso2009datacenter}. So the workloads we want to characterize should be recently used and implemented with emerging techniques in different domains.
\subsection{Can one data analysis workload represent all?} \label{why_not}

In the rest of this paper, in comparison with other benchmarks, we find that data analysis workloads
and scale-out service workloads have some common characteristics, e.g, notable frond-end stalls, which is consist with CloudSuite.
However, CloudSuite is biased towards service workloads, and they choose four service
workloads among the six benchmarks and only include one data analysis
workload: \emph{Naive Bayes}. Can \emph{Naive Bayes} workload represent all of the data analysis workloads? The answer
is no.
%Through our experiments, we find that different data analysis applications not only have different speed up performance but also have different micro-architecture level characteristics (Section \ref{characterization_BDbenchmarks}).

%The scale-out ability is the most popular user observation metric in the date center computing, which measures the speed up
%performance of the application through node scale increasing.

\begin{figure}
\centering
\includegraphics[scale=0.5]{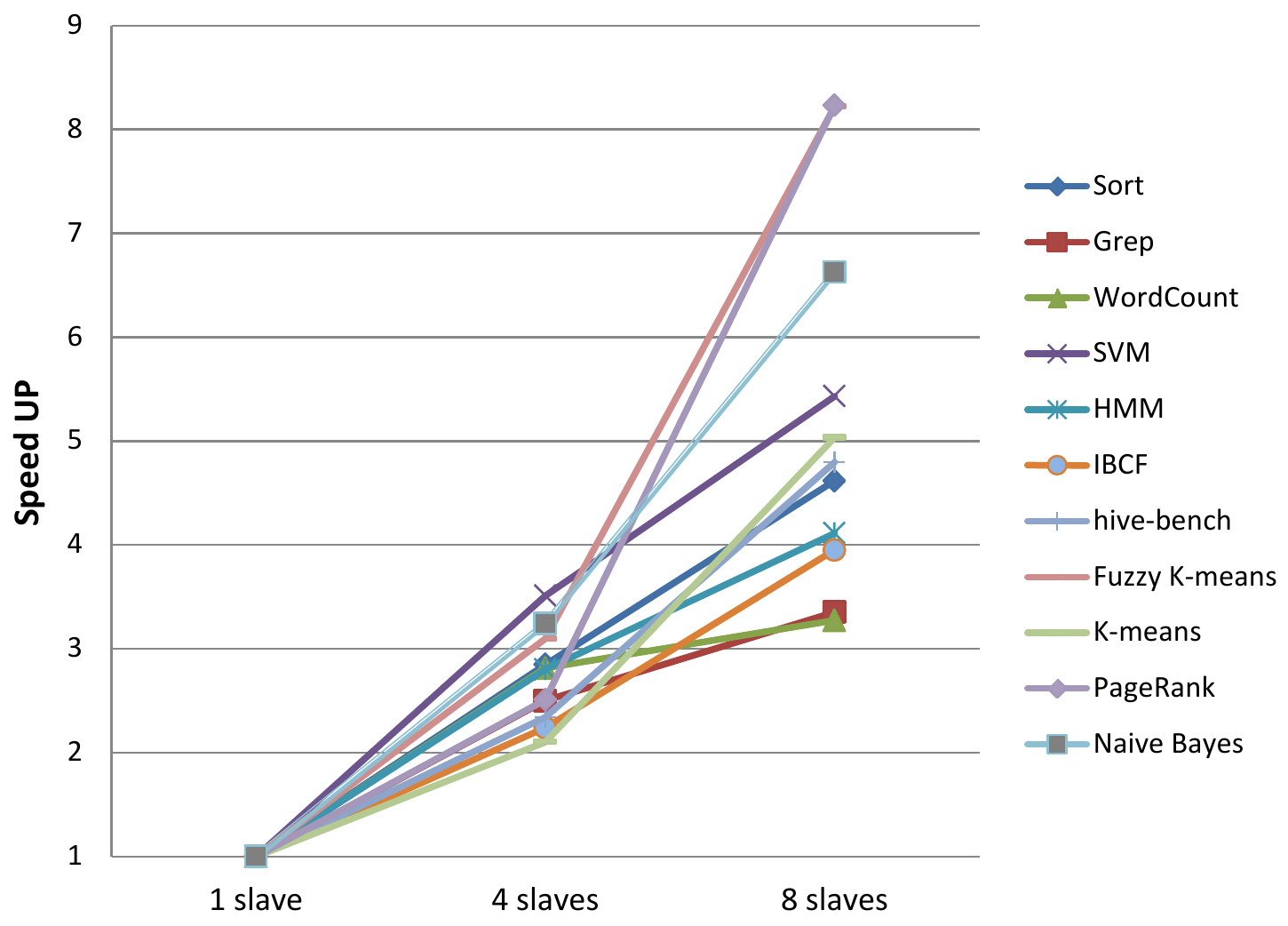}
\caption{Varied speed up performance of eleven data analysis workloads. Only Bayes is included into CloudSuite.}\label{speedup}
\end{figure}

In this subsection, we perform the  experiments on a 9-node Hadoop cluster, including one master node. The configuration of the system and the Hadoop environment
is the same as that in Section \ref{character}. For the conciseness,
we leave the configuration details in Section \ref{character}. % in  which only include one master node, the nodes in our Hadoop cluster are connected through 1 Gb ethernet network. Each node's configuration listed in Section \ref{hwcon}.% of two- Intel Xeon E5645 (Westmere) processors and 32 GB memory.
We choose eleven representative workloads
including \emph{Naive Bayes}, the details of which can be found at Section \ref{workloads_choice}, and use the same input data sets (listed in Table \ref{workloads}).
%and Hadoop configuration to drive them.
We change the number of slave nodes  from 1 to 8, using the run time on the one-slave system as as a baseline performance data, and  then calculate the speed
up.
 %through dividing one-slave execution time by multi-slaves execution time.
  As shown in Figure. \ref{speedup}, for different workloads, the speed up data on an eight-node system  range from 3.3 to 8.2, and the value of \emph{Naive Bayes} is 6.6,  which indicates that the data analysis workloads are diverse in terms of performance characteristics.
%  , implied that these applications have different
%user observation characteristics. In our experiments, some applications' speed up at 8 nodes are greater than 8, this is
%because that we use the same data set for all of the experiments, which resulted in the data set(around 150GB ) maybe too large for
%one slaves to process.
 Meanwhile our  experiments in Section \ref{characterization_BDbenchmarks} also show that different data analysis applications have
different micro-architecture level characteristics.

In a word, only one application is not enough to represent various categories of data analysis  workloads.
%, so we choose eleven representative
%workloads.

%Different applications have
%different demands on resources, some is memory sensitive, whereas some is CPU sensitive.
%Different application also have
%, among six  benchmarks,
%including  data analytic, data serving,
%media streaming, web search, software testing, web serving.
%a scale-out cloud computing benchmark
%suite, including six kinds of benchmarks, data analytic, data serving,
%media streaming, web Search, software testing, web serving. CloudSuite only has one
%application in data analysis benchmark, {\em classification}, which is
%not enough to represent various categories data analysis workloads.
%Our experiments in Section \ref{characterization_BDbenchmarks}
%also shows that different data analysis applications
%have different micro-architecture level %scale-out performance
%characteristics. %and a single application cannot capture the key
%characteristics of all big data analysis workloads.

\subsection{Workloads Choosing}  \label{workloads_choice}
In order to find workloads which meet all the requirements listed in Section \ref{reforus}, we firstly decide and rank the main application domains according to a widely
acceptable metric---the number of pageviews and daily visitors,
and then single out the main applications from the most important application domains.
We investigate the top sites listed in Alexa \cite{Alexa},
of which the rank of sites is calculated using a combination of average daily visitors and page views. %over the past month.
We classified the top 20 sites into 5 categories including search
engine, social network, electronic commerce, media streaming and others.
Figure \ref{share} shows the categories and their respective share. To keep concise and simple, we focus on
the top three application domains: \emph{search engine}, \emph{social networks} and \emph{electronic commerce}.

We choose the most popular applications in those three application domains. Table \ref{algorithm_scenarios} shows application scenarios of each workload, which is characterized in this paper, indicating most of our chosen workloads are intersections among three domains.

\subsubsection{Basic operations}
\emph{Sort}, \emph{WordCount} and \emph{Grep} are three basic operations which are frequently used in data analysis fields. %those three application domain.
%very basic operations, frequently used
%Such as the sort is used in sorting the match
%items of search results; word count is used when search engine process the indexing and
%grep is used in match the goods in electronic commerce.
%They are
Several
researchers also use them in their work~\cite{dean2008mapreduce}.
%Three of the applications in our benchmark suite can be seen as basic operations, namely Sort, Word Count and Grep.
%In our paper, they are implemented
%with the MapReduce framework.
\emph{Sort} ranks the records by their key.
\emph{WordCount} reads text files and counts the number of occurrences of each word.
\emph{Grep} extracts matching strings from text files and counts the number of the
occurrence of the matching strings.
%We got those applications from Hadoop examples.

\subsubsection{Classification and Clustering}
The classification and clustering algorithms are
also widely used in those three application domains. The goods information in electronic commerce and key words of search engines are the input data of
classification and clustering applications.
We choose \emph{Naive Bayes} and \emph{Support Vector Machine (in short, SVM)} from numerous classification algorithms.
\emph{Naive Bayes} is a simple probabilistic classifier which applies Bayes'
theorem
with strong (naive) independence assumptions. %We got it from Mahout \cite{mahout}.
\emph{SVM}, a very widely used classification algorithm, maps the
training samples as
points in a high dimension space so that the samples of the separate categories
can be divided by a clear gap. And new samples are categorized based on the side of
the gap they fall on.
%Many benchmarks in different fields include SVM as one of their workloads for its
%popularity\cite{narayanan2006minebench} \cite{clemons2011mevbench}.
%So we would like choose a distributed version SVM in our paper. %benchmark suites.
%For we do not find a open source implementation, we implemented it using hadoop.

%Naive Bayes, Support Vector Machine (SVM), K-means, Item Based Collaborative
%Filtering (IBCF), Frequent Pattern Growth (FPG), Hidden Markov Model (HMM),
%Kernel Principal Component Analysis (KPCA) ~\cite{wu2008top}.

%\textbf{Clustering}
For the clustering applications,
we choose %the distributed implementation of the
\emph{K-means} and \emph{Fuzzy K-means} algorithms
%into our
%benchmark suite as some data mining benchmarks do \cite{narayanan2006minebench}
because
those two clustering algorithms are so famous in many application scenarios.
\emph{K-means } partitions observations into a fixed
number clusters in which each observation belongs to the (only) cluster with the
nearest mean.
\emph{Fuzzy K-means} is an extension of \emph{K-means}, but quite different from the latter.
It is statistically
formalized and allows soft clusters where an observation can belong to multiple
clusters with given probabilities. %We have tow versions of K-means and fuzzy K-means
%implementations. One is hadoop the other is MPI.
%The reason why We include both MPI version and hadoop version is that the programming
%models have a big impaction on the OS-level and architecture=level metrics.
%The MPI version applications we got from Mahout \cite{mahout} while for the MPI
%version we got them from IBM Parallel Machine Learning \cite{pml}.

\subsubsection{Recommendation}
The recommendation algorithms are widely used  to recommend goods, friends, and key words in electronic commerce,
social network and  search engine, respectively.
We choose an \emph{Item Based Collaborative Filtering (in short IBCF)} application. It
estimates a user's preference towards an item by looking at his/her preferences
towards related  items. %It is widely used by Internet service companies for finding clients' interests and recommending appropriate products/services to users.
%We got it from Mahout \cite{mahout}.

%\textbf{Relationship Mining}
%
%We chose FPG (Frequent Pattern Growth), a well known association rule mining algorithm, as the
%relationship mining specific implementation.
%%FPG is employed in many areas such as market basket analysis, web usage mining, and
%%intrusion detection.
%FPG discovers meaningful relations between variables in
%large databases by adopting a divide-and-conquer approach to decompose both the
%mining tasks and the databases. We got it from Mahout \cite{mahout}.

\subsubsection{Segmentation}
Segmentation is very important for web search, especially for a language like Chinese.
We implement a segmentation algorithm
using the \emph{Hidden Markov Model (in short, HMM)}.
%The Hidden Markov Model is widely used in temporal pattern recognition such as speech,
%handwriting, gesture recognition, part-of-speech tagging, and
%bio-informatics.
\emph{HMM} is a statistical Markov model in which the
system being modeled is assumed to be a Markov process with unobserved hidden
states.

%\textbf{Dimension reduction}
%Here we include PCA (Principal Component Analysis) and
%KPCA (Kernel Principal Component Analysis) into our benchmark suites.
%PCA is a mathematical procedure that uses an orthogonal transformation to convert a set of
%observations of possibly correlated variables into a set of values of linearly uncorrelated
%variables called principal components.
%KPCA is an extension of principal component analysis (PCA) using techniques of kernel methods.
%Using a kernel, the originally linear operations of PCA are done in a reproducing
%kernel Hilbert space with a non-linear mapping.
%Kernel principal component analysis is a powerful technique for
%extracting a structure from potentially high-dimensional data sets and has been
%applied to in fields like face recognition \cite{kim2002face}.
%We got both of them from IBM Parallel Machine Learning \cite{pml}.

%{\em \bf ZLX: state which benchmark comes from where?}  For the aforementioned
%algorithms, we select the implementations from ProfSearch~\cite{scisearch},
%MapReduce based Mahout~\cite{mahout}, MPI based IBM Parallel Machine Learning
%(PML) tool~\cite{pml}.  In particular, xxx and xxx are from ProfSearch. xxx are
%from Mahout. xxx are from PML.  For K-means, we have included a MapReduce
%version from xxx and an MPI version from xxx.

%\textbf{PageRank}

\subsubsection{Graph calculation}
%We use graph500\cite{graph} as the graph calculation workloads. It implement depth-first search in graph.
%PageRank is a very important workloads in search engine.
\emph{PageRank} \cite{page1999pagerank} is a typical graph calculation algorithm on link analysis,  and \emph{PageRank-like} algorithms are frequently used in search engine.
%It is a link analysis algorithm, named after Larry Page.%, one of the founders of Google, and used by the Google web search engine, that assigns a numerical weighting to each entry of a hyperlinked set of documents.
%We got it %the applications, which implement the PageRank algorithm,
%from Mahout\cite{mahout}.
%
%\textbf{Vector calculation}
%
%For the vector calculation applications, we implemented it using AllPairs \cite{moretti2010all},
%an abstraction for data-intensive computing,which can perform parallel computing on multiple node.
%This application calculates tow papers' similarity through a TF-IDF model which is widely
%used in information retrieval and text mining.

%The workloads we choosing in this paper are really frequently used in real world,
%Those workloads
%covers mostly used algorithms in data mining\cite{wu2008top}, and can
%be used in many application scenarios. Table \ref{algorithm_scenarios} lists some
%application scenarios of those workloads.

\subsubsection{Data warehouse operations}

%Data warehouse operations are important workloads in data centers.
In the three application domains we mentioned above, data warehouse are often used to manage data.
The previously published Hive-bench~\cite{hive-bench} is a benchmark for the
data warehouse operations based on the Hive \cite{hive}. %implementation of Hadoop~.
Here, we include a series of representative SQL-like
statements in \emph{Hive-bench} as a part of our workloads.

%\subsection{Input data sets}
%
%Input data sets are an integral part of a benchmark suite. The data sets for the
%DataAnalysisBench benchmark suite are either unmodified data obtained from
%real-world deployments, e.g., genome data for SAND and traffic accident data for
%frequent pattern growth (FPG), or synthetic data generated according to given
%models, e.g., the input document set for Word Count.
%
%Table~\ref{workloads} gives detail information of each benchmark in our
%benchmark suite and its data input set.
\begin{table*}[hbtp]
\caption{Representative Data Analysis Workloads}\label{workloads}
\centering
\begin{tabular}{|c|c|c|c|c|} \hline
No. & Workload & Input Data Size    & \#Retired Instructions (Billions)  & Source \\ \hline
 1 & Sort      & 150 GB  documents & 4578 &Hadoop example \\ \hline
 2 & WordCount  & 154 GB documents  & 3533 &Hadoop example\\ \hline
 3 & Grep   & 154 GB  documents & 1499 & Hadoop example \\ \hline
 4 & Naive Bayes  & 147 GB text  & 68131& mahout\cite{mahout} \\ \hline
 5 & SVM  & 148 GB html file    & 2051 &our implementation\\   \hline
 %6 & K-means  &MapReduce  &text document from sougou \cite{sogoukmeans} & 39974 &data mining  \\ \hline
 6 & K-means  & 150 GB vector  & 3227 &mahout\\ \hline
 % 7& fuzzy K-means &MapReduce &text document from sougou \cite{sogoukmeans} & 145&data mining\\ \hline
 7 & Fuzzy K-means & 150 GB vector &15470 & mahout  \\ \hline
 8 & IBCF  & 147 GB ratings data & 32340 & mahout \\  \hline
% 9& FPG &MapReduce & 1.4 GB traffic accident data \cite{FPG}    & 25&Relationship Mining\\ \hline
  9& HMM   & 147 GB html file  &1841  &our implementation \\  \hline
 10 & PageRank  & 187 GB web page & 18470  &mahout\\  \hline
 11 & Hive-bench & 156 GB  DBtable  & 3659 &Hivebench\\ \hline
\end{tabular}
\end{table*}
%To summarize, DataAnalysisBench meets all the requirements listed in Section \ref{requirements}:
%\begin{itemize}
%  \item Eighteen representative data analysis applications employing state-of-art algorithms are included into our DataAnalysisBench. We will depict those applications in section Section \ref{app_selected}, and show why we choose those applications for our benchmark suite.
%  \item DataAnalysisBench contains applications written in three different programming models: MapReduce, MPI, and WorkQueue.
%  \item All of DataAnalysisBench workloads are distributed. Those workloads can be easily deployed  on a data center, which has a large amount of nodes.
%\end{itemize}
%are presented in Table \ref{workloads}.
%The details information of DataAnalysisBench can be found in .
%Because of the large computational cost and our testbed's scale, we chose a medium input data set for Section \ref{character}'s experiments.

\begin{table}[hbtp]
\caption{Scenarios of Data Analysis.}\label{algorithm_scenarios}
\centering
\begin{tabular}{|p{2cm}|p{3.0cm}|p{2.5cm}|} \hline
  Name& Domain &Scenarios \\ \hline
          &search engine     &  	Log analysis \\ \cline{3-3}
 	Grep    & social network  &Web information extraction \\ \cline{3-3}
 	        & electronic commerce  &Fuzzy search  \\ \hline
  Bayes & social network  & Spam recognition\\ \cline{3-3}
             & electronic commerce	& Web page classification \\ \hline
           & social network      & Image Processing \\ \cline{3-3}
 	SVM       & electronic commerce   & Data Mining \\ \cline{3-3}
 	          &      & Text Categorization \\ \hline
  PageRank &search engine & Compute the page rank \\ \hline
  Fuzzy  &search engine    & Image processing \\ \cline{3-3}
  K-means& social network &High-resolution landform\\
  K-means        & electronic commerce & classification \\ \hline %\cline{2-2}
         % &     & Speech recognition \\ \hline
         & social network   &  	Speech recognition \\ \cline{3-3}
   HMM   & search engine  & 	Word Segmentation \\ \cline{3-3}
          &  & 	Handwriting recognition \\ \hline
             &search engine   & Word frequency count \\ \cline{3-3}
  WordCount & social network   & Calculating the TF-IDF value \\ \cline{3-3}
 	             & electronic commerce &Obtaining the user operations count\\ \hline
%  Vector calculation & Similarity calculation \\ \cline{2-2}
%    & Bioinformatics \\ \hline
  Sort & electronic commerce & Document sorting\\ \cline{3-3}
	     & search engine & Pages sorting \\
       & social network &  \\ \hline
 %  & computer vision \\ \cline{2-2}
%dimension reduction &pattern recognition \\ \cline{2-2}
%  &face recognition \\ \hline
\end{tabular}
\end{table}

\section{Experimental Setup}
\label{character}

This section firstly describes the experimental environments on which we conduct our study,
and then explains our experiment methodology.

%This section first introduces our testbed and experimental methodology; and then
%characterizes the data analysis workloads we chosen and compares them with
%traditional benchmarks from SPECCPU 2006, HPCC, and SPECWeb 2005.
%Also we compare them with CloudSuite\cite{ferdman2011clearing},
%which is a scale-out benchmark suite for cloud computing.

%; and then
%shows the effect of programming models.

%The goal of this section is three-fold. First, it provides quantitative proof
%that places data center applications into a workload class different from
%desktop, parallel, and traditional server workloads.  Second, it illustrates the
%distinct performance characteristics of each of the common data center
%domains.
%\includegraphics[scale=0.5]{test.jpg}

%\subsection{Experimental setups}

% for getting data from hardware  performance counters.
\subsection{Hardware Configurations} \label{hwcon}

We use a 5-node Hadoop cluster (one master and four slaves) to run all data analysis workloads.
%We benchmark one of the slave node of the hadoop cluster, which running the workloads.
The nodes in our Hadoop cluster are connected through 1 Gb ethernet network.  Each
node has two Intel Xeon E5645 (Westmere) processors and 32 GB memory.
A Xeon E5645 processor includes six physical  out-of-order cores with
speculative pipelines. Each core has private L1 and L2
caches, and all cores share the L3 cache. Table~\ref{hwconfigeration}
lists the important hardware configurations of the processor.

\begin{table}
\caption{Details of Hardware Configurations.}\label{hwconfigeration}
\center
\begin{tabular}{|c|c|}
  \hline
  %\multicolumn{2}{|c|}{CPU Type} & \multicolumn{2}{|c|}{Intel CPU Core} \\ \hline
  %\multicolumn{2}{|c|}{Intel \textregistered Xeon}  &\multicolumn{2}{|c|}{4cores@1.6G} \\  \hline
  CPU Type & Intel \textregistered Xeon E5645\\ \hline
  \# Cores & 6 cores@2.4G \\ \hline
  \# threads& 12 threads \\ \hline
	\#Sockets & 2 \\ \hline
  \hline
  ITLB & 4-way set associative, 64 entries \\ \hline
  DTLB & 4-way set associative, 64 entries \\ \hline
  L2 TLB& 4-way associative, 512 entries \\ \hline
  L1 DCache & 32KB, 8-way associative, 64 byte/line \\ \hline
  L1 ICache & 32KB, 4-way associative, 64 byte/line \\ \hline
  L2 Cache & 256 KB, 8-way associative, 64 byte/line \\ \hline
  L3 Cache &  12 MB, 16-way associative, 64 byte/line \\ \hline
  Memory & 32 GB , DDR3 \\  \hline
\end{tabular}
\end{table}

\subsection{Hadoop Cluster Environments} \label{configuration}
All the workloads are implemented on the  Hadoop system,
 which is an open source MapReduce implementation. %, base applications for the Map-reduce programming model
%is so famous in big data processing fields.
The version of Hadoop and JDK is 1.0.2 and 1.6.0, respectively.
For data warehouse workloads, we use Hive of the 0.6 version.
Each node runs Linux CentOS 5.5  with the 2.6.34 Linux kernel.
Each slave node is configured with 24 map task slots
and 12 reduce task slots. %One map
%task slot is assigned per hardware thread and one reduce task slot is assigned per core, and
For each map and reduce task, we assigned  1 GB Java heap in order to achieve better performance.

\subsection{Compared Benchmarks setups} \label{traditionalben}
In addition to data analysis workloads, we
deployed several
 benchmark suites, including SPEC CPU2006, HPCC, and SPECweb 2005, CloudSuite---a scale-out benchmark suite for cloud computing \cite{ferdman2011clearing}, and compared them with data analysis workloads.
%We also compared them with CloudSuite,
%which is

%For all the applications are MapReduce based workloads , we use Hadoop
%1.0.2 and JDK 1.6.
%For MPI based benchmarks, we use MPICH2 version 1.3.1 and
%gcc 4.1.2.  For Hive based workload, we use Hadoop 1.0.2 and Hive 0.6.
%For Rubis benchmark,
%it has three different parts (mysql, jboss and apache). So we use show the mean of performance
%metrics.
%The WorkQueue based workload is also running on an eight-node testbed.
\subsubsection{Traditional benchmarks setups}
SPEC CPU2006: we run the official applications with the first reference input, reporting results averaged into two groups, integer benchmarks (\emph{SPECINT}) and floating point benchmarks (\emph{SPECFP}). The gcc which we used to compile the SPEC CPU is version 4.1.2.

%benchmarks with their reference input data sets on a single core.
	HPCC: we deploy HPCC --a representative HPC benchmark suite. The HPCC version is 1.4. It has seven benchmarks\footnote{\emph{HPL} solves linear  equations. \emph{STREAM} is a simple synthetic benchmark, streaming access memory. \emph{RandomAccess}  updates (remote) memory randomly. \emph{DGEMM} performs matrix multiplications.
\emph{FFT} performs discrete fourier transform. \emph{COMM} is a set of tests to measure latency and bandwidth of the interconnection system.}, including \emph{HPL}, \emph{STREAM}, \emph{PTRANS}, \emph{RandomAccess}, \emph{DGEMM}, \emph{FFT}, and \emph{COMM}.
We %separated the sequential running form and
run each benchmark respectively.
	%\item Linpack: We run the Linpack, which is a representative workload in HPC(hight performance computing) field, on one node.
  %\item
	
	SPECweb 2005: we run the bank application as the Web server on one node with 24 GB data set.
	%and the input data size  is 24 GB.
	We use distributed clients to generate the workloads, and the number of the total simultaneous sessions is 3000.
 % We run the e-banking workloads on one node.
  % \item CloudSuite: We run all the CloudSuite benchmark and show the arithmetic mean.
%\end{itemize}

\subsubsection{CloudSuite Setups}

CloudSuite has six benchmarks, including one data analysis workload---
\emph{Naive Bayes}. We also choose \emph{Naive Bayes} as one of the
representative data analysis workloads with a larger data input set (147 GB). In \cite{ferdman2011clearing}, the data input size is only 4.5 GB.

We set up the other five  benchmarks following the introduction on the CloudSuite web site \cite{cloudsuite}.
		
		Data Serving: we benchmark \emph{Cassandra} 0.7.3 database with 30 million records. The request is generated by a YCSB \cite{cooper2010benchmarking} client with a 50:50 ratio of read to update.
    %\item
		
		Media Streaming: we use \emph{Darwin} streaming server 6.0.6. We set 20 Java processes and issue 20 client threads by using the Faban driver \cite{faban} with GetMediumLow 70 and GetshortHi 30.
		
		Software Testing: we use the \emph{cloud9} execution engine, and run the printf.bc coreutils binary file.
		
		Web Search: we benchmark a distributed  \emph{Nutch} 1.1 index server. The index and data segment size is 17, and 35 GB, respectively.
		
		Web Serving: we characterize a front end of \emph{Olio} server. We simulate 500 concurrent users to send requests with 30 seconds ramp-up time and  300 seconds steady state time.

\subsubsection{Data Analysis Workload Setups}
Table~\ref{workloads} presents the size of input data set and the  instructions retired of each data analysis workload. The input data size varies from 147 to 187 GB. In comparison with  that of CloudSuite described in \cite{ferdman2011clearing}, our approach are more pragmatic, and we deploy a larger data input that are stored in both memory and disk systems instead of completely  storing data (only  4.5 GB for \emph{Naive Bayes} in \cite{ferdman2011clearing}) in memory. The number of instructions retired of the data analysis workloads
ranges from thousand of billions to tens of thousands of billions, which indicate that those applications
are not trivial ones.

%\subsection{Performance Data collection}
\subsection{Experimental Methodology} \label{eMethodology}
%The general computing pattern in modern computing systems is data movement and calculating.
%It is to say sending the correct data to computing units.
%So we focus on the event which could delay sending data to computing unit or stall the calculating
%pipeline.
%There are many factors that can affect the performance either in data movement or calculating.
%For examples, a cache miss or TLB miss may delay the data movement; a misprediction of branch
%instructions will cause a pipeline flush further result in penalties.
%The detailed accounting for how the
%CPU cycles are used is useful to analyze applications' performance. But the retirement
%centric  analysis is difficulty to do this when the target architecture
%has Out of Order (OoO)
%execution because the entire objective of
Modern superscalar Out-of-Order (OoO) processors prevent us from breaking down the execution time precisely
due to overlapped work in the pipeline \cite{ferdman2011clearing}
\cite{keeton1998performance} \cite{eyerman2006performance}.
The retirement centric analysis is also difficult to
account how the CPU cycles are used because the pipelines will continue executing instructions even though the
instruction retirement is blocked \cite{levinthal18027cycle}.
So in this paper we focus on counting  %present results based on the performance counters that reflect
\emph{cycles stalled due to resource conflict}, e.g. the
reorder buffer full stall, which prevents new instructions from entering the pipelines.

We get the micro-architectural data by using hardware performance counters to measure the architectural events. In order to monitor micro-architectural events, a Xeon processor provides several performance event
select MSRs (Model Specific Registers), which specify hardware events to be counted, and performance monitoring counter MSRs, which store results of performance monitoring events.
We use Perf---a profiling tool for Linux 2.6+ based systems \cite{perf}, to manipulate those MSRs by specifying the event numbers and corresponding unit masks.
%For the limited number of the hardware performance counters (performance event select MSRs),
We  collect about 20 events whose number and corresponding unit masks can be found in the Intel Developer's Manual \cite{intelref}.
In addition, we access the proc file system to collect OS-level
performance data, such as the number of disk writes. %RSS (Resident Set Size), and others.

%Map-reduce jobs have
%several stages, map, shuffle and reduce. Different applications have different
%percentage of map and reduce stages. % The map tasks and reduce tasks can be
%overlapping running at the same time.
%The overlapping time is also different according to
%specific application.
%For a job, only collecting data for a short period can not reflect the whole program's performance. So we
%benchmark the whole run time of the application.

We perform a ramp-up period for each application, and then start collecting the
performance data.
Different from the experiment methodology of CloudSuite, which only performs 180-second measurement, the performance data we collected cover the whole lifetime of each application, including map, shuffle, and reduce stages.
We collect the data of all the four working nodes and report the
mean value.

%\subsection{Input Data Sets}
%The applications in data centers always process large amount of data. The data scale can
%range from Giga byte to Peta byte, which far exceeds the memory capacity of data centers.
%In order to imitate the realistic environment,
%On our Hadoop cluster
%(4 working nodes), we use a larger input data set, exceeding the total memory amount of the four working nodes.
%The exact input data size of each workload is listed in Table \ref{workloads}.
%The amount of input data set exceeds the total mount of working nodes' memory.
%the memory can not hold all the data set. Each major page fault will cause tens nanosecond penalty to access disk.
%So in this paper we choose a larger input data set, , which is more closer to realistic environment with our 4 working slave node scale cluster.

%Second, for each workload, we collect the performance data of the whole lifetime in stead of a sample of a short period, e.g., 180 seconds.
%, which is affect by experiment environments
%such as the data set, storage (RAM disk v.s. disk) and operating system version etc.

\section{Results}
\label{characterization_BDbenchmarks}

%First, we give out an overview of experiment results, and then

We provide a detailed analysis of
the inefficiencies of running data analysis workloads on modern OoO
processors in the rest of this section.

\subsection{Instructions Execution}

\begin{figure}
\centering
\includegraphics[scale=0.4]{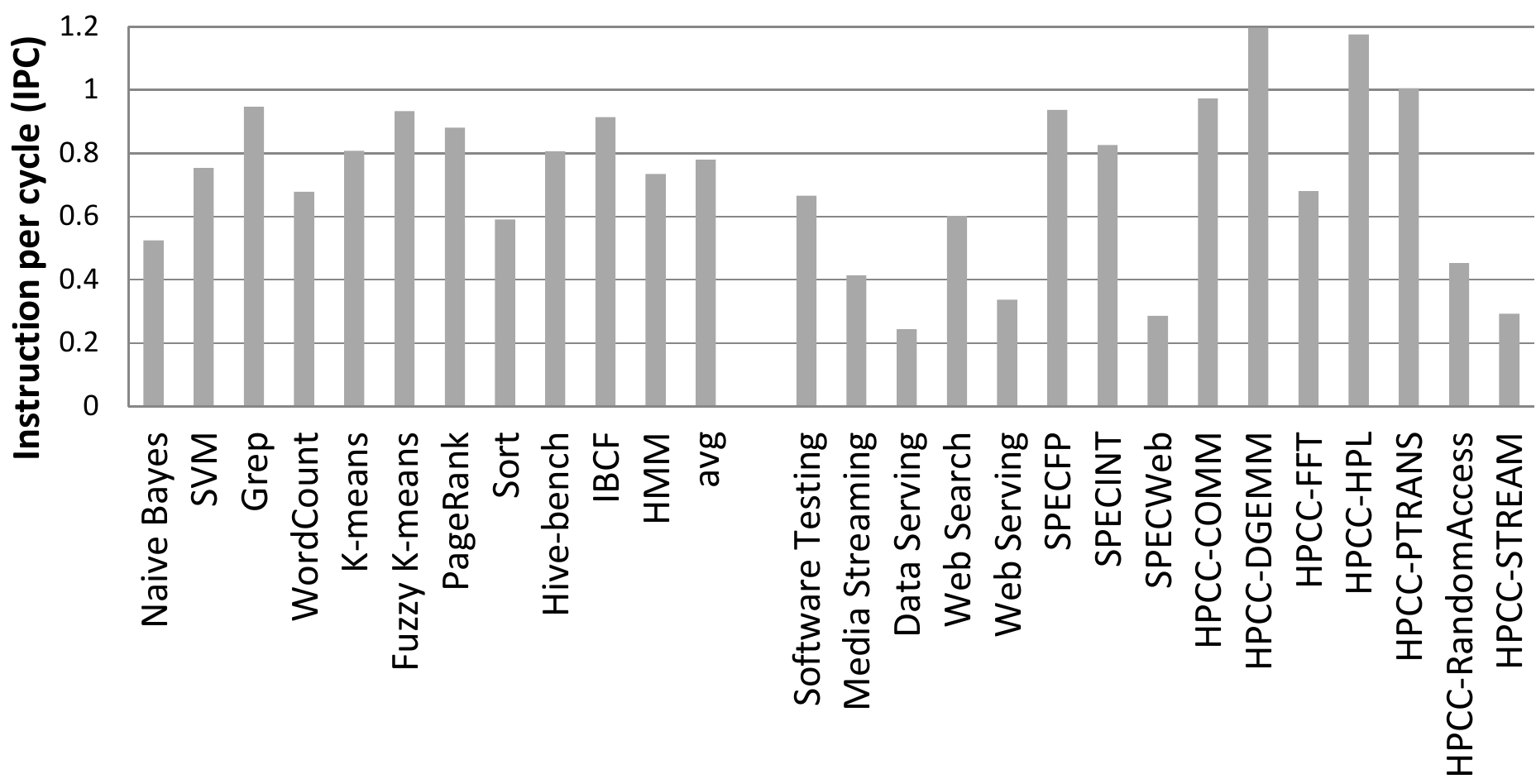}
\caption{Instructions per cycle for each workload.}\label{ipc}
\end{figure}

%We can compute the IPC of each big data analysis workloads with the event that count
%the number of cycles and the number of instruction retired.

Instruction per cycle (in short IPC) is used to measure instruction level parallelism,
indicating how many instructions can execute simultaneously.
Our processors have 6 cores,  and each core can commit up to 4 instructions on each
cycle in theory.
However, for different workloads, IPC can be limited by pipeline stall and
data or  instructions dependencies.
% among instructions and data.

Figure \ref{ipc} shows IPC  of each workload.
The CloudSuite has six benchmarks, among which we report the Naive Bayes on the leftmost side, separated from the other five workloads (in the middle side), since \emph{Naive Bayes} is also included into our eleven workloads.
%and we list five of them in the right side of the figure because we have already
%included the Naive Bayes workload as one of our big data analysis workloads, which listed
%at the leftmost of the figure.

The main workloads of CloudSuite (four among six) are service workloads: \emph{Media Streaming, Data Services, Web Services, and Web Search}. From Figure \ref{ipc}, we can observe that service workloads, including four of CloudSuite and \emph{SPECweb} has the lower IPC (all less than 0.6) in comparison with
the other workloads, including our chosen data analysis workloads, \emph{SPECFP}, \emph{SPECINT}, and most of HPCC workloads.

Most of data analysis workloads have middle-level IPC values, greater than that of the service workloads.
The IPC of the eleven data analysis workloads ranges from 0.52 to 0.95 with an average value of 0.78. The avg bar in Figure \ref{ipc} means the average IPC of the eleven data analysis workloads. \emph{Naive Bayes} has the  lowest IPC value among the eleven data analysis workloads.
The IPCs of the HPCC workloads have a large discrepancy
among each workload since they are all micro-benchmark designed for measuring different aspects of systems. For example, \emph{HPCC-HPL} and \emph{HPCC-DGEMM} are computation-intensive, and hence have a higher IPC (close to 1.2). While  \emph{HPCC-STREAM} are designed to stream access memory,  it has poor temporal locality, causing long-latency memory accesses, and hence it has lower IPCs (less than 0.5).

%The low IPC can be explain by the pipeline stall and data or instruction dependency.
%Such as,
%Naive Bayes, data serving and web serving owns the lowest IPC mainly
%for their high
%L2 cache misses (as shown in Figure \ref{cache_miss}), which cause long latency data access.
%Sort has low IPC caused mainly by the inefficacy front end which
%has more instruction cache miss (Figure \ref{L1cache}) resulting in pipeline
%front end stalls (Figure \ref{stall}).
%On the average the big data analysis applications yield medium level of IPC which is better
%than data serving, media streaming and etc; worse than web search, HPL
%(High Performance Linpack) in HPCC and etc.

%greater than 1 such as SPEC int, which is consists with \cite{ferdman2011clearing}.
%TPC-C has
%the highest IPC, which is because it has few cache misses as shown in Section
%\ref{mhb}. This also indicates that it has better instruction-level
%parallelism.
\begin{figure}
\centering
\includegraphics[scale=0.4]{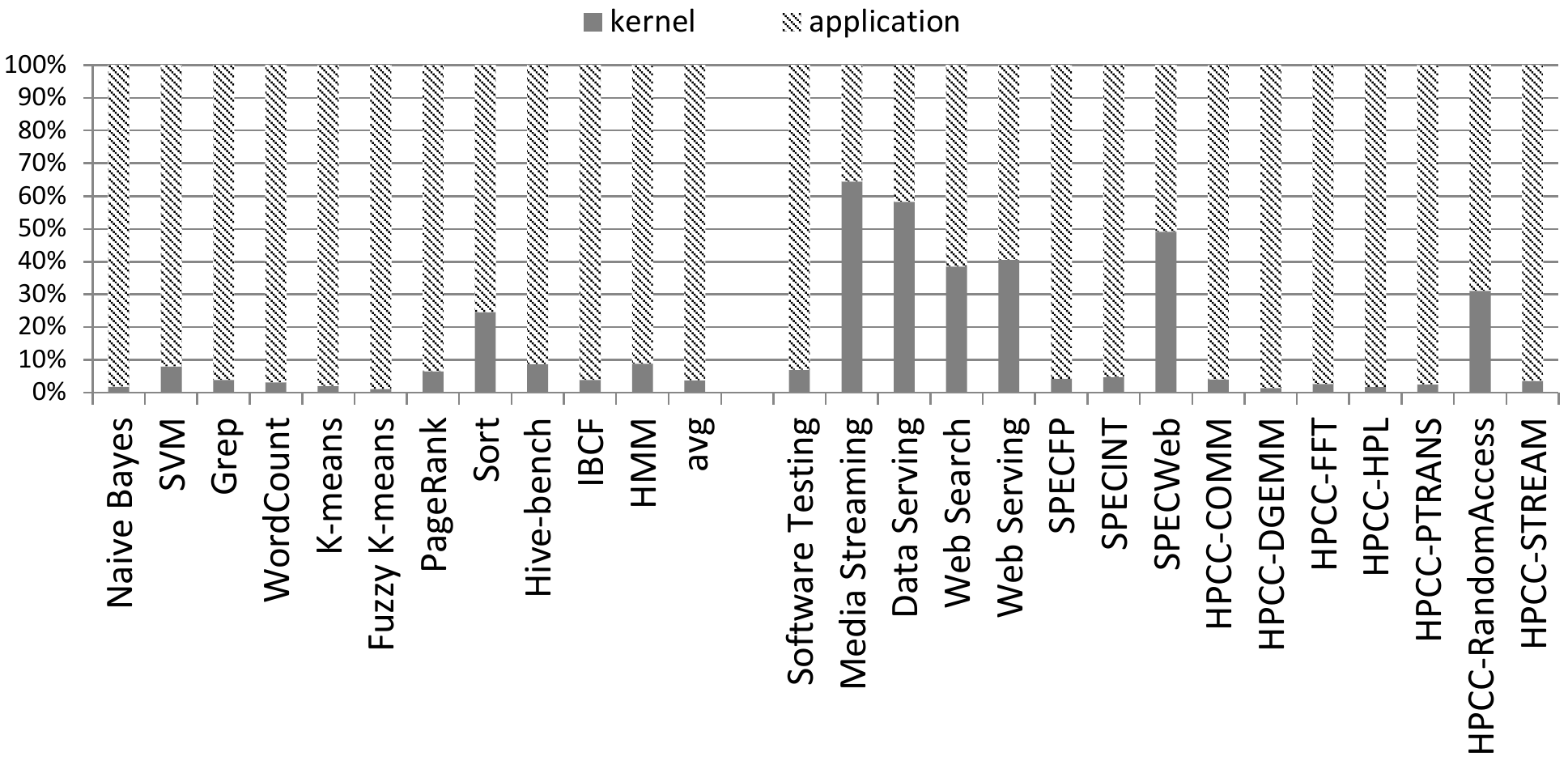}
\caption{User and Kernel Instructions Breakdown.}\label{ins_share}
\end{figure}

Figure \ref{ins_share} illustrates the retired instructions breakdown of each workload.
We also notice that the service workloads (four of CloudSuite and \emph{SPECweb}) execute a large percentage of kernel-mode instructions (greater than 40\%), while most data analysis workloads execute a small percentage of kernel-mode instructions.
The service workloads have higher percentages of kernel-mode instructions because serving a large amount of requests will result in a large number of network and disk activities.
%and may induce long latency to  further lower IPC.

%We can find that different from the main workloads of CloudSuite, including media streaming, data serving, web serving, and the traditional server workload---SEPCWeb and etc,
%the big data analysis
%applications execute more user mode (application) instruction than kernel's.

%One of the main reason which caused high percentage of kernel mode instruction execution
%is that the sort have more frequently disk write than other applications.

\begin{figure}
\centering
\includegraphics[scale=0.5]{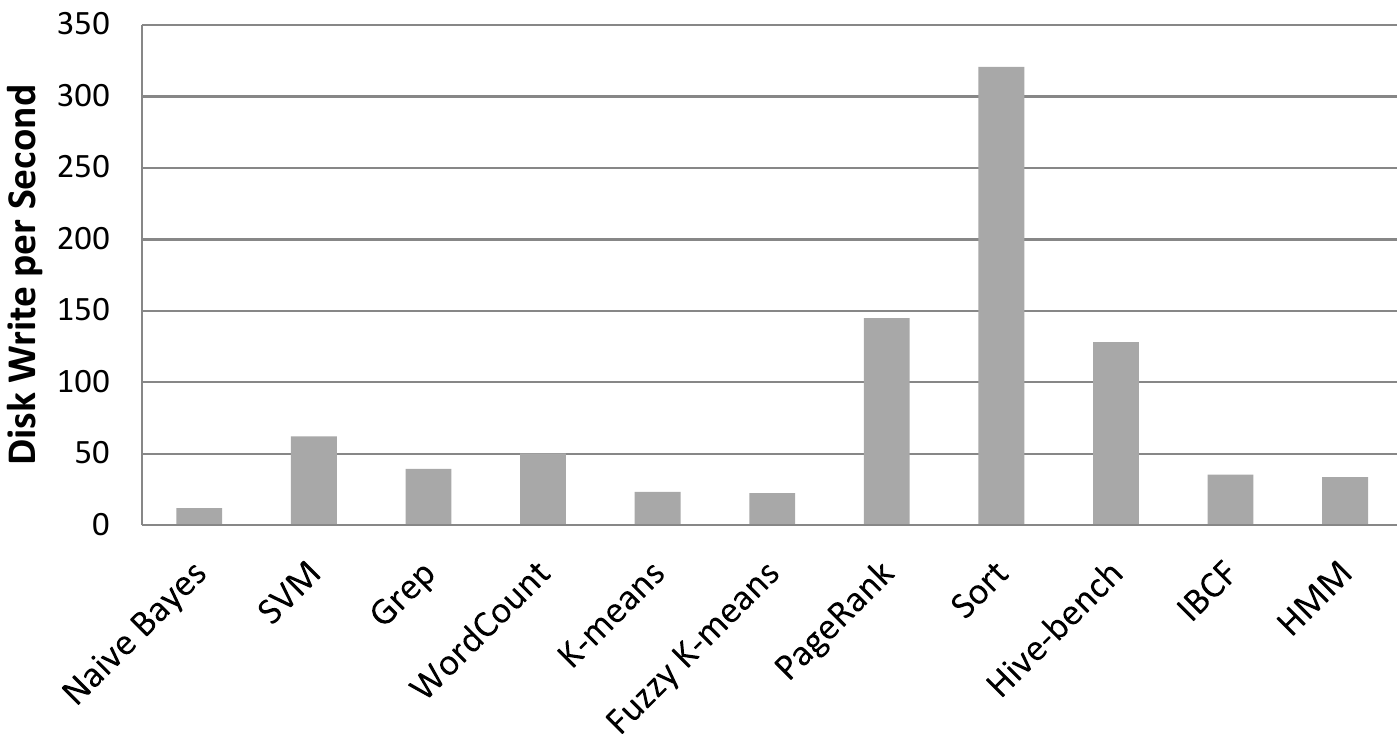}
%\caption{Frequently disk writes cause high percentage of kernel mode instruction executions.}\label{diskwrite}
\caption{Disk Writes per Second.}\label{diskwrite}

\end{figure}

Among the data analysis workloads,  only \emph{Sort} has a high proportion (about 24\%) of kernel-mode instructions whereas on average the data analysis workloads only have about 4\% instructions executed in kernel-mode.
This is caused by the two unique characteristics of \emph{Sort}.
The first one is that different from most of the data analysis workloads,
the input data size of \emph{Sort} is equal to the output data size.
So in each stage of the MapReduce job, the system will write large amount of output data to local disk or transfer a large amount of data over network.  This characteristic makes Sort have more I/O operations than other workloads.
The second unique characteristic is that \emph{Sort} has simple computing logic, only comparing. So it can process a large amount data in a short time.
Those characteristics let \emph{Sort}  involve more frequently I/O operations (disk and network).
%process a large amount of data in a short time, and hence
%This algorithm only sort the input data and do not filter or
%extract data from the input.
%The amount of intermediate data
%\footnote{The map-reduce jobs have several stages, each stages the job will
%write the intermediate data in to local disk.}
%is also equal to the input data.
%The intermediate data and output data the sort write to disk is more than other big data
%analysis applications.
So in comparison with other data analysis workloads, \emph{Sort} are more OS-intensive.
Figure \ref{diskwrite} depicts disk writes per second of each data analysis workload.
We can find that \emph{Sort} has the highest disk writes frequency.
%The network communication package transferred frequency for data analysis workloads has the similar result to  disk writes frequency in Figure \ref{diskwrite} (we do not show the figure for the limited space).
We also observed that network communication activities of  \emph{Sort} are also more frequent than that of the other data analysis workloads.

Among the HPCC workloads, \emph{RandomAccess} has a large percentage of kernel-mode instructions (about 31\%). \emph{RandomAccess} measures the rate of integer random updates of (remote) memory.
 An update is a read-modify-write operation on a table of 64-bit words, and it involves a large amount of  $copy\_user\_generic\_string$ system calls.
 The other factors contributing a large percentage of kernel-mode instructions need further investigations.
% which is a user space memory copy function.
%There must be some other factors that affect the instruction breakdown.
%instruction execution percentage between kernel mode and application mode,
%But we do not further discussed, for it exceeds the scope of this paper.}

\textbf{\emph{Observations}}:

Data analysis workloads have higher IPC than that of services workloads, which are characterized by CloudSuite and traditional web server workloads, e.g., \emph{SPECweb2005},  while lower than that of computation-intensive workloads, e.g., \emph{HPC-HPL, HPC-DGEMM}.
Meanwhile we also observe that the most of data analysis workloads involve less kernel-mode instructions than that of the services workloads.
%So the system engineers should not pay much attention
%to kernel code efficiency if their big data applications do not frequently I/O operation like \emph{Sort}.
%while more than that of the HPCC and SPEC CPU workloads.
%, \emph{SPECFP}, and \emph{SPECINT} workloads.
%This observation indicates that data analysis workloads  would be more suited by
%architectures offering a modest degree of superscalar out-of-order execution  than that of service workloads  \cite{ferdman2011clearing}.
%In addition, data analysis workloads do not heavily rely upon OS functions like the service workloads, and they may benefit from a light-weight OS, which is practiced in HPC communities.

%There must be some other factors that affect the instruction execution percentage between
%kernel mode and user mode, but we do not further discussed, for it is outside the scope
%of this paper.

\subsection{Pipeline Behaviors}\label{rrs}
Modern processor implements dynamic execution using out of order and speculative engine.
The whole processor architecture can be divided into two parts: including an in-order front
end, which fetches, decodes, and issues instructions, and an out-of-order back end, which
executes instructions and write data back to register files.
A stall can happen in any part of the pipeline.
In this paper we focus on the major pipeline stalls (\emph{not exhausted}),   including front end (instruction fetch), register
allocation table (in short \emph{RAT}), load-store buffers, reservation station (in short \emph{RS}), and re-order buffer (in short \emph{ROB}).
For modern X86 architecture, front end will fetch instructions from L1 Instruction cache and then decode the CISC
instructions into RISC-like instructions, which  Intel calls micro-operation.
\emph{RAT} will change the registers used by the program
into internal registers available. %, allowing the micro-operation
%to run at the same time of
%another micro-operation that uses the exact same standard register.
Load-store buffers are also known as memory order buffers, holding in-flight memory
micro-operations (load and store), and they ensure that writes to memory take place in the
right order.
\emph{RS} queues micro-operations
until all source operands are ready.
\emph{ROB}  track all micro-operations
in-flight and make the out-of-order executed
instructions retire in order.

\begin{figure}
\centering
\includegraphics[scale=0.45]{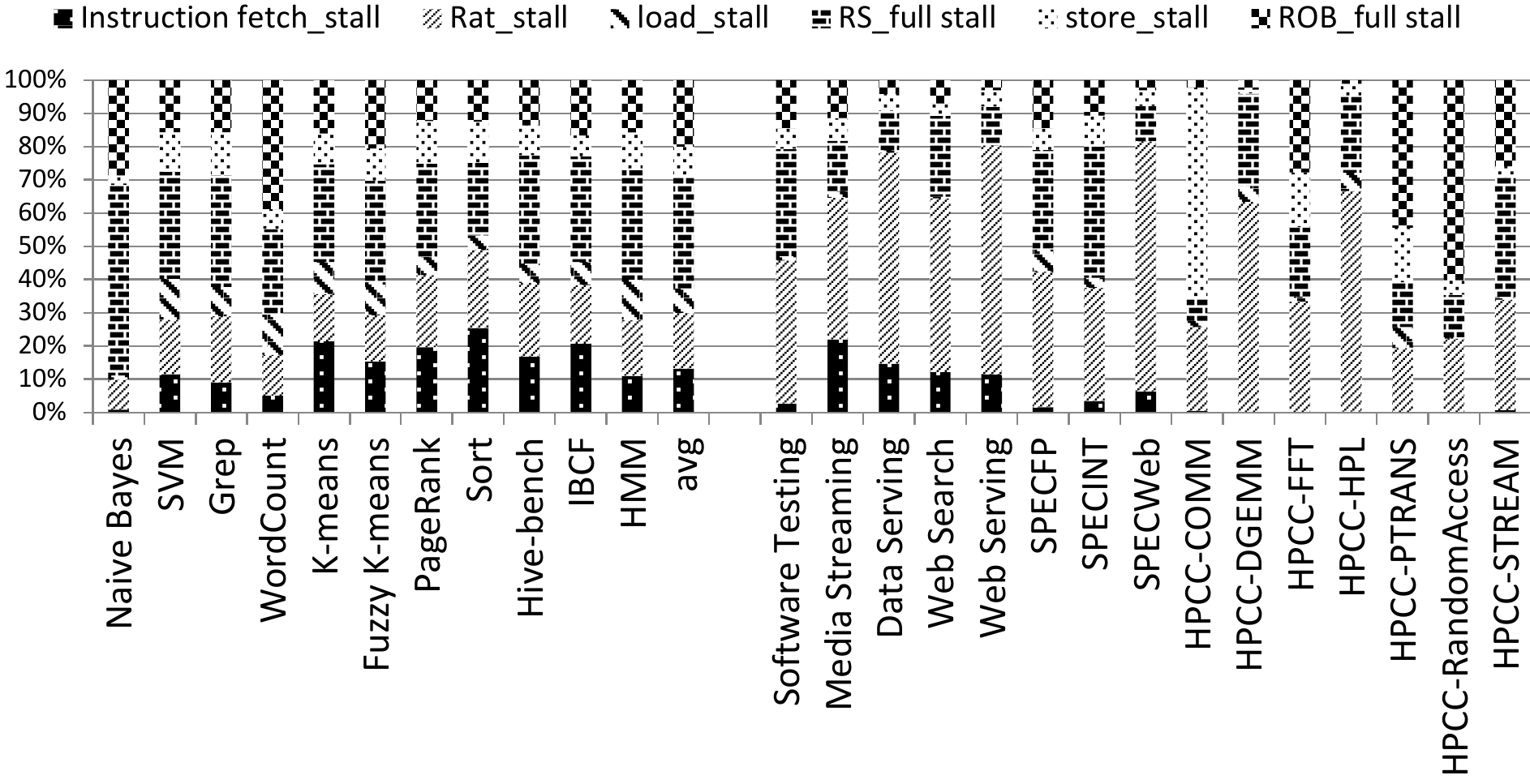}
\caption{Pipeline Stall Break Down of Each Workload}\label{stall}
\end{figure}

Figure \ref{stall} presents those major stalls in pipelines for each workload including instruction fetch stalls,
RAT stalls, load buffer full stalls, store buffer full stalls,
RS full stalls, and ROB full stalls.
We can get the blocked cycles of those kind of stalls mentioned above by using hardware performance
counters.
Different kinds of pipeline stalls may occur simultaneously, that is to say, the stall
cycles may overlap.
For example,  when the back end is stalled due to \emph{RS} full, the front end can also be stalled due to L1 instruction cache misses.
%The back end stalled cycles and front end stalled cycles can overlap.}
So in Figure \ref{stall}, we report the normalized values of the stalled cycles.
We calculate the normalized value by using the following way:
we sum up all the blocked cycles for all kinds of stalls as the total blocked cycles.
Then we divide each kind of stall's blocked cycles by the total blocked cycles as their percentage in Figure \ref{stall}.

Different from HPCC and SPEC CPU2006 workloads,
the data analysis workloads and service workloads suffer from notable front end stalls,
which are mainly caused by
L1 instruction cache miss, ITLB (Instruction Translation Lookaside Buffer)
miss or ITLB fault, reported in front-end performance data  in Section \ref{front} .
The notable front end stalls indicates the instruction fetch
unit inefficiency. Our observation corroborates  the previous work \cite{ferdman2011clearing}.
The front end inefficiency may caused by high-level languages and third-party libraries
used by the data analysis and service workloads. The complicated framework and middleware will increase the binary size of the whole application even though they only implement a simple algorithm.
%\emph{\textbf{possible reasons!}}
%The service workloads including media streaming, data severing, web search, and web severing also own
%a notable front end stall, but they suffer
%Nearly for all the workloads suffer RAT stalls.
%But we find the right side benchmarks own

We also find that there are notable differences in terms of stalls breakdown between the data analysis workloads and service workloads (including four service workloads of CloudSuite and SPECWeb).
The latter workloads own a large percentage of RAT stalls, which may be caused by partial register stall or
register read port conflicts. While the data analysis workloads suffer from more RS stalls and ROB stalls, which are caused by
limited RS and ROB entries.
\emph{RAT} and instruction fetch stalls occur before instruction entering the out-of-order part of the pipeline while the RS and ROB stalls occur at the out of order part of the pipeline.
The service workloads (including \emph{Media Streaming, Data Severing, Web Severing,
Web Search and SPECweb}) have 60\% RAT
stalls and 13\% instruction fetch stalls on average, whereas the data analysis workloads have about 37\% RS full stalls and 20\% ROB full stalls on average.
So we can find that the data analysis workloads suffer more stalls in the out-of-order part of the pipeline, while the service workloads suffers more stalls in the in-order part of the pipeline.
Further investigation
is necessary to understand the root cause behind the differences between two kinds of workloads.

%pipercentage stalls.

For the HPCC workloads are composed of micro benchmarks and kernel programs, different programs focus on a specific
aspect of the system. So their stall data vary dramatically from each other in Figure \ref{stall}.
% when compared with
%real applications.
%The workloads in HPCC can be seen as micro benchmarks, they focus on a specific
%aspect of the system. Such as the HPL focus on the floating point instruction execution
%ability, so it make data more fit into the cache line and use mostly floating point
%operations.
%The HPCC steam workload owns poor temporal locality, once the data it accesses, then it will
%hardly access again.
%Those stalls seem meaningless for
%The real applications seldom do like those.
%So the pipeline stalls here may be feel useless.
%Although there are more physical registers available than instruction
%set architecture (ISA)
%defined logical registers, the register re-allocation (tracked in the Register Allocation Table) still becomes
%a bottleneck.

%Our initial guess is  that the integer register files
%are overwhelmed by the predominate integer instructions while the floating point register
%files mostly stay idle.
%Although the stall cycles may have overlap, the total stall cycles in the figure
%can also reflect the efficiency of the core. The stalls, which mostly caused by
%dependency, prevent the instruction retired from pipeline until it gets the required
%resources. The more pipeline stall the less instructions retired.
%We can find that if the workload owns a high
%pipeline stall it has a poor IPC in figure \ref{ipc}.

\emph{\textbf{Implications:}}

Corroborating previous work \cite{ferdman2011clearing}, both the data analysis workloads and the service workloads suffer from notable front-end stalls (i.e. instruction fetch stall).
The instruction fetch stalls means that the front end has to wait for fetching instructions,
which may be caused by two factors: deep memory hierarchy with long latency in modern processor \cite{ferdman2011clearing}, and  large binary size complicated by  high-level language and third-party libraries.
%The large amount of instruction fetch stalls may cause less instructions send to back end.
%Even though modern processor have many execution units, they can not be used efficiently. So more small cores may more efficiency as \cite{janapa2010web} said.}

However, we note the significant differences between the data analysis workloads and the service workloads in terms of stall breakdown:
the data analysis workloads suffer more stalls in the out-of-order
part of the pipeline,
while the service workloads suffer
more stalls before instructions entering the out-of-order part.
This observation can give us some implications
about how to alleviate the bottlenecks in pipeline,
although one well known consequence is that right after of alleviating the bottleneck, the next bottleneck emerges \cite{thomadakis2011architecture}.
%For data analysis workloads, the architecture should pay more attention to alleviate the our-of-order stalls, such as alleviating data dependency.
%For service workloads, the architecture should pay more attention to the in-order part, such
%as alleviating the register port conflicts.

%This observation indicates the service workloads will not benefit from aggressive or even modest  degree of superscalar out-of-order execution.
%For data analysis applications, more instruction fetch stalls and the out-order execution stalls means that the front end has to wait for the instruction coming and back end has to wait for solving dependency. Although modern processor have many execution units, they can not be used efficiently. So more small cores may more efficiency as \cite{janapa2010web} said. %Also for the programs, a batch of small jobs with small binary size and less data dependency can achieve better performance than large jobs.

\subsection{Front-end Behaviors} \label{front}
\begin{figure}
\centering
\includegraphics[scale=0.45]{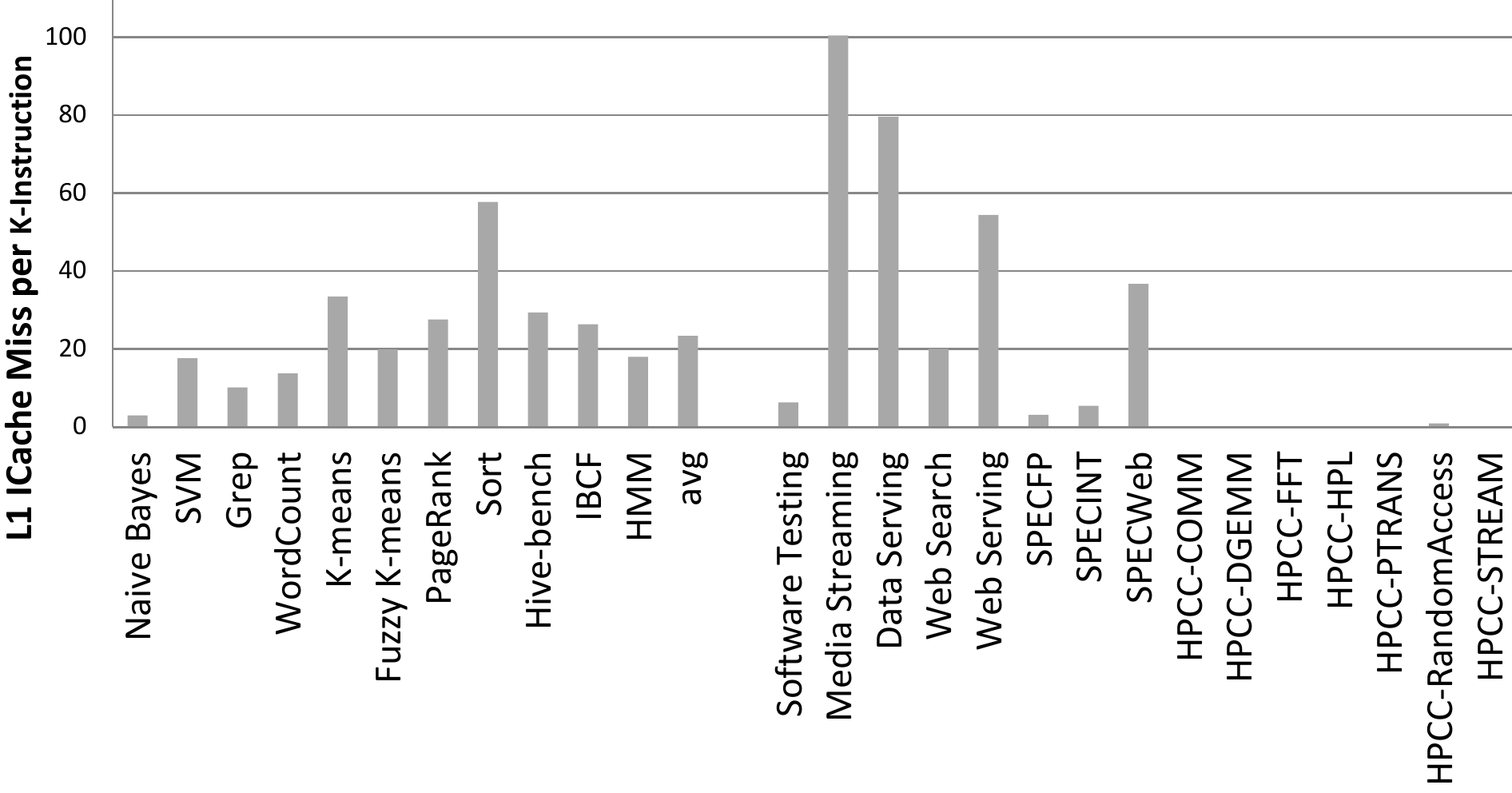}
\caption{L1 Instruction Cache misses per thousand instructions.}\label{L1cache}
\end{figure}

%Front-end is the place where instructions enter the pipeline.
The
instruction-fetch stall will prevent core from
making forward progress due to lack of instructions.
Instruction cache and instruction Translation Look-aside Buffer (TLB) %, which will be discussed in Section \ref{page_walk},
are two fundamental components, which  must be accessed when fetching instructions from memory.
Instruction cache is the place where the fetch unit
directly get instructions.
TLB stores page table entries (PTE),
which are used to translate virtual addresses to physical addresses.
%Each time a virtual memory access, the processor searches the
%TLB for the virtual page number of the page that is being accessed,
If a TLB entry is found with a matching virtual page
number, a TLB hit occurs and the processor can use the retrieved physical address
to access memory.
Otherwise there is a TLB miss, the processor has to look up the page table, which called a page walk.
The page walk is an expensive operation.
%With a three-level page table, three memory
%accesses would be required. In other words, it would result in
%four physical memory accesses.}
%So the L1 instruction cache miss and instruction TLB
%fault will cause at least several cycles' latency of instruction fetch.

\begin{figure}
\centering
\includegraphics[scale=0.45]{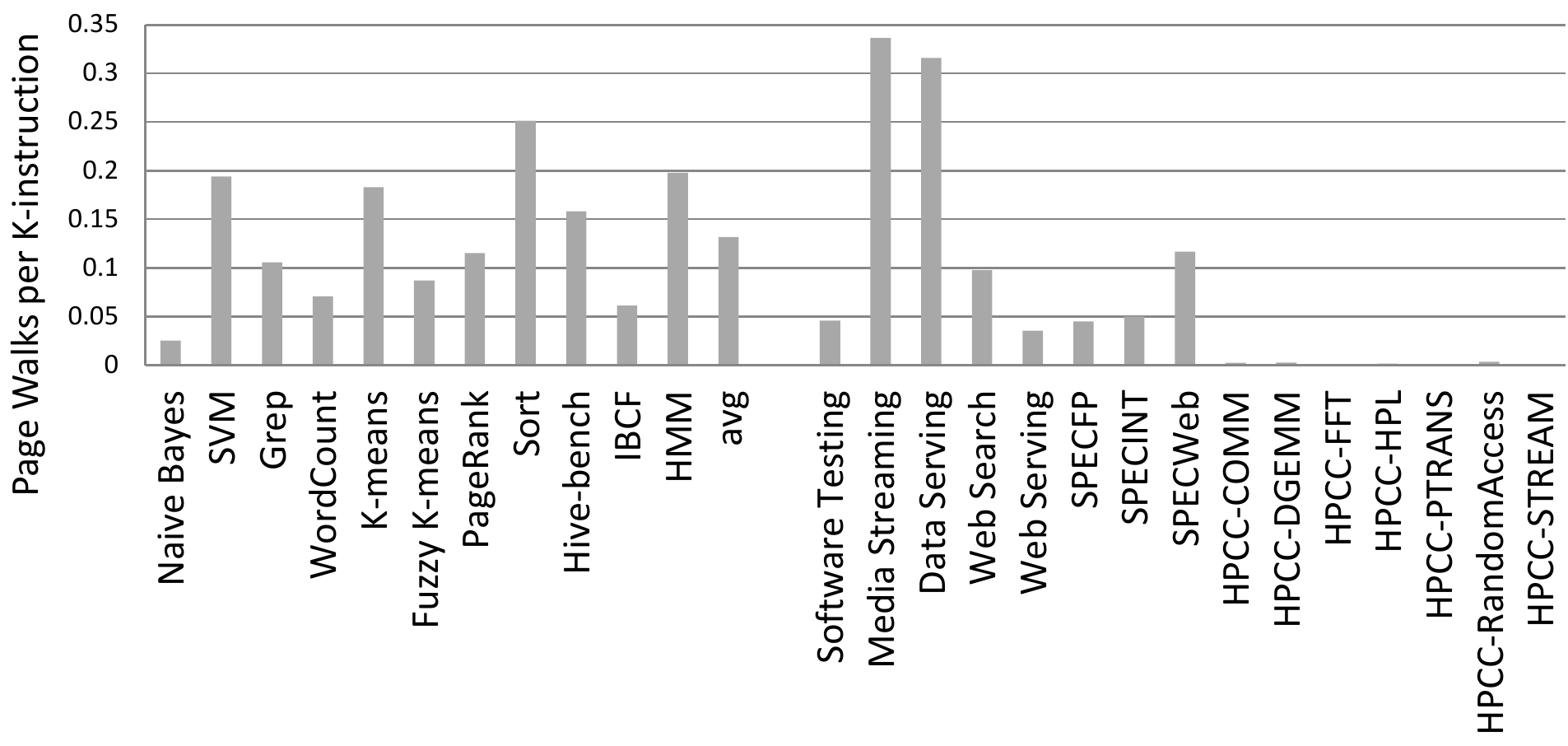}
\caption{ITLB miss caused completed page walks per thousand instructions.}\label{itlb}
\end{figure}

Figure \ref{L1cache} and Figure \ref{itlb} present the L1 instruction cache misses and the completed page walks caused by instruction TLB misses per thousand instructions, respectively.
On average, the data analysis workloads generate about 23 L1
instruction cache misses per thousand instructions.
They own higher L1 instruction cache misses
than that of  \emph{SPECINT}, \emph{SPECFP}, and all the HPCC workloads.
Most of the data analysis applications have less L1 instruction cache misses than that of the service workloads including \emph{Media Streaming, Data Severing, Web Serving} and \emph{SPECweb}.
\emph{Media streaming} has a larger instruction footprint and  suffers from
severe L1 Instruction cache misses, whose L1 Instruction cache misses are
about three times more than the average of that of the data analysis workloads.
Higher L1 instruction cache misses result in higher instruction fetch stalls as shown in
Figure \ref{stall}, indicating less efficiency of the front-end.
For most of the others benchmarks, the L1 instruction cache misses are really very rare, especially the
HPCC workloads, whose instruction footprint is relatively small.
%Similar to big data analysis workloads,
%data serving workloads owns a higher instruction cache misses.
%That means they own a large instruction set, which cause the instruction cache
%change content frequently.

%This indicate that it own a large instruction set, which cause the instruction cache change content frequently.

%\subsection{TLB Behaviors} \label{page_walk}
%Modern processor accesses virtual memory not only to extend primary memory, but also to make such an extension as easy as possible for programmers to use. A virtual address must be translated to the corresponding physical address before the processor can get data or instructions.

Consistent with the performance trend of L1 instruction cache misses, the data analysis workloads' completed page walks caused by instruction TLB miss are more frequently than that of \emph{SPECINT}, \emph{SPECFP}, and all HPCC workloads.
Some service workloads (\emph{Media Streaming} and \emph{Data Serving} workloads) have more
completed page walks than that of the data analysis workloads.
Page walks will cause a long latency instruction fetch stall, waiting for correct physical addresses so as to fetch instructions, and  hence result in inefficiency of front end.
Among the data analysis workloads, \emph{Naive Bayes} is an exception  with the smallest L1 instruction cache misses and completed page walks caused by instruction TLB misses, so it can  not represent the spectrum of all data analysis workloads.

%, and hence it can not represent the full spectrum of the data analysis workloads.

\emph{\textbf{Implications:}}

Improving the L1 instruction cache and instruction TLB hit ratios can improve the performance of data analysis workloads, especially the service workloads.
%the data center workloads (both data analysis workloads and service workloads). %, especially for the service workloads.
%by reducing the latency that pipeline back end waiting for instructions coming.
%So the L1 instruction cache and instruction TLB performance become focus.}
The third-party libraries and high-level languages used by datacenter workloads may enlarge the
binary size of applications and further aggravate the inefficiency of instruction cache and TLB.
So when
writing the program (with the support of third-party libraries and high-level languages), the engineers should pay more attention to the code size.

% quality. Some compiler optimization can increases the occurrence of the instructions executed in parallel, minimize the branch penalty, but increase program code
%size, which will enlarge the instruction footprint and further
%cause the instruction cache and TLB less efficiency. The
%system engineers need to decide whether those kind of optimization can improve the whole performance of data analysis applications.
%compared with
%other applications.

\subsection{Unified Cache and Data TLB Behaviors} \label{mhb}
%Memory hierarchy is a significant bottleneck in modern computing systems.
The manufacturers of processors introduce a deep memory hierarchy to reduce the
performance impacts of memory wall.
Modern processors own three-level caches.
%High level cache owns small capacity and less access latency, the lower level cache has
%large capacity but more access latency.
 A miss  penalty of last-level cache can reach
up to several hundred cycles in modern processors.

\begin{figure}
\centering
\includegraphics[scale=0.45]{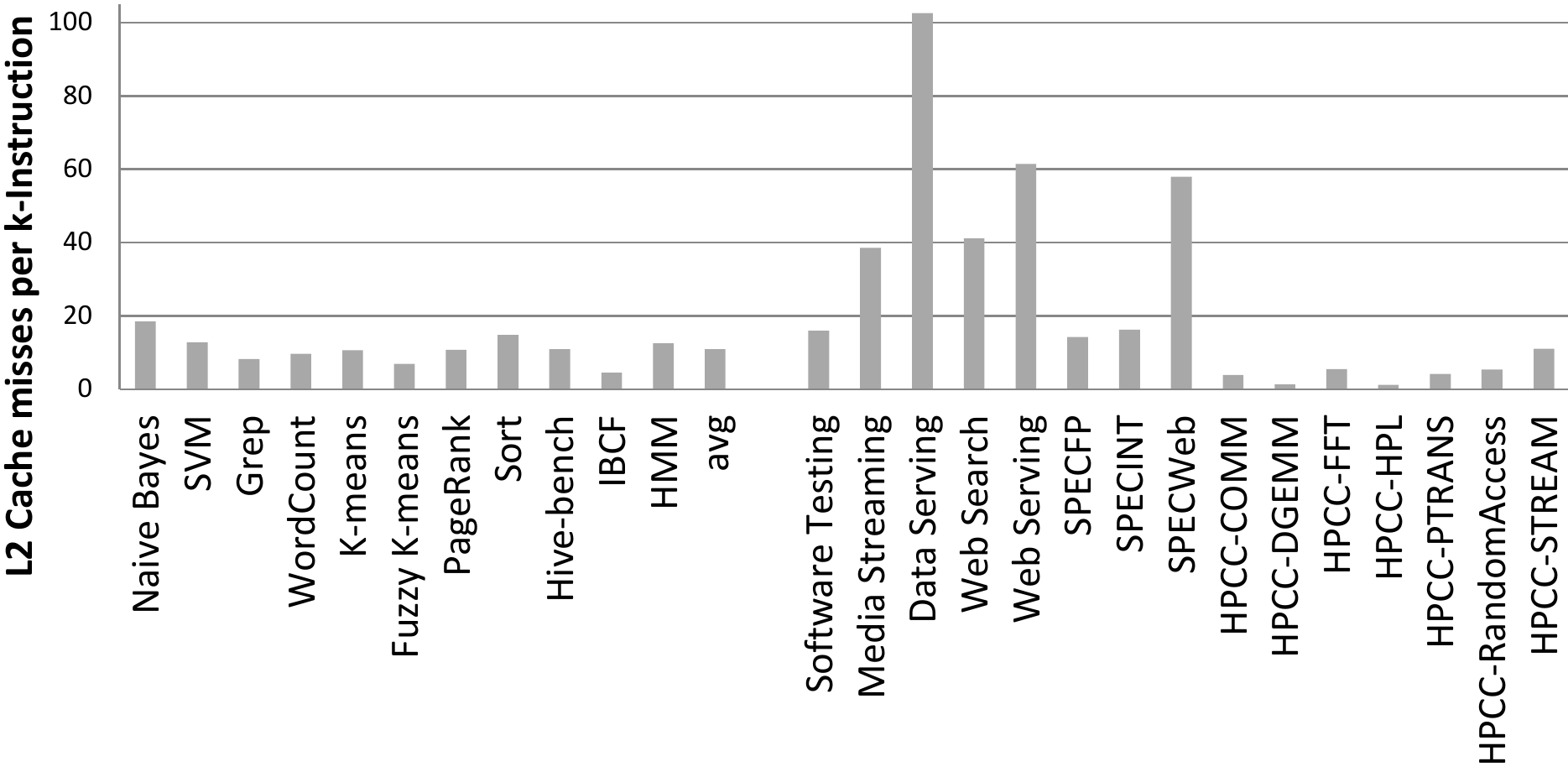}
\caption{L2 cache misses per thousand instructions.}\label{cache_miss}
\end{figure}

\begin{figure}
\centering
\includegraphics[scale=0.45]{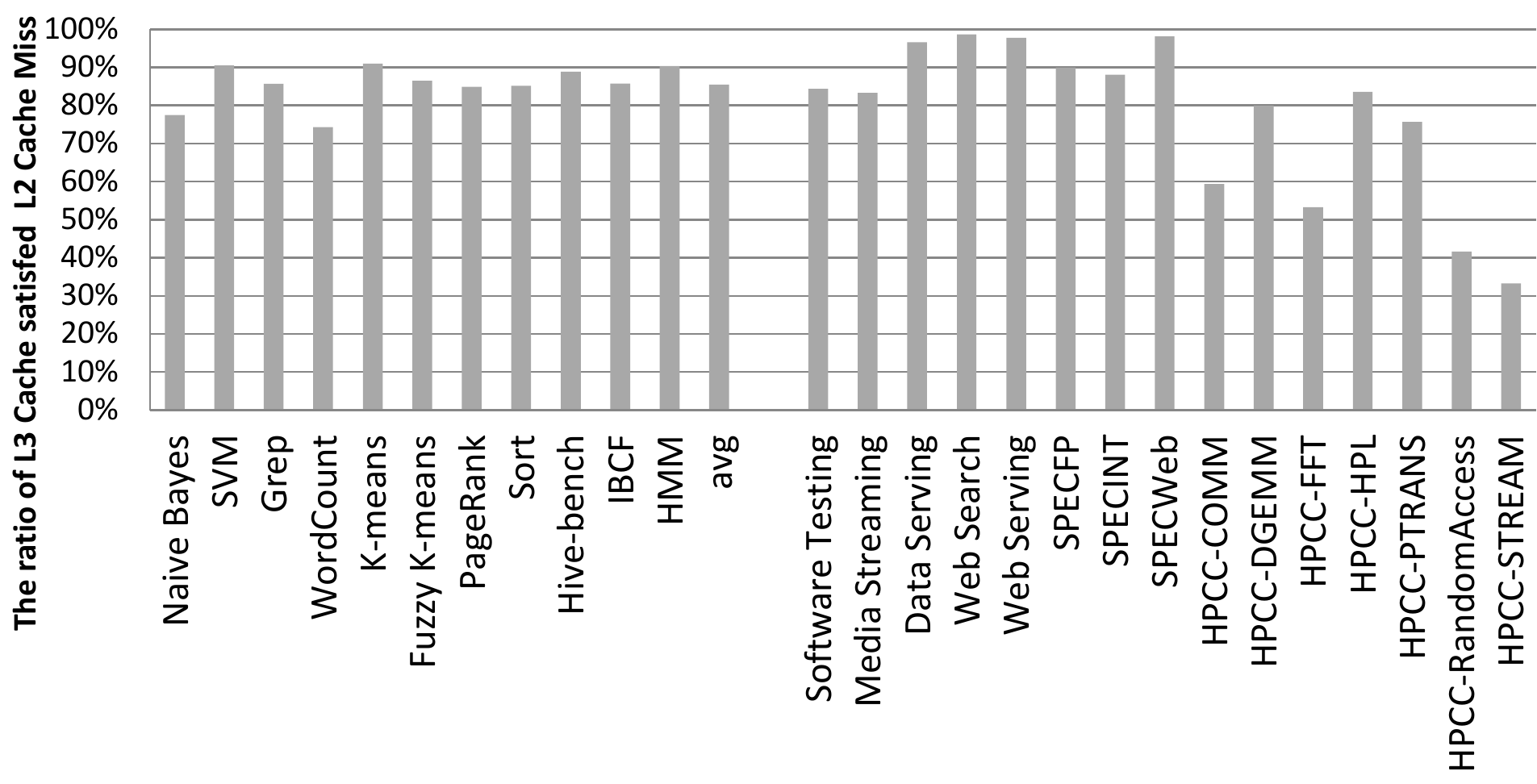}
\caption{The ratio of L3 cache satisfying L2 cache misses.}\label{l3cachel2}
\end{figure}

Figure \ref{cache_miss} shows the L2 cache misses per thousand instructions.  Figure \ref{l3cachel2} reports the ratio of L3 cache hits over L2 cache misses.
This ratio can be calculated by using  Equation~\ref{eq:equ1}.
Please note that we do not analyze the L1 data cache statistics for the miss penalty can be hidden by
the out-of-order cores \cite{karkhanis2004first}.    % and have little impacts on performance.
\begin{equation} \label{eq:equ1}
  ratio =\frac{L2 ~ cache ~misses- L3 ~cache ~misses}{L2 ~cache~ misses}
\end{equation}

For most of the data analysis workloads, they have lower  L2 cache misses  (about 11 L2 cache misses per thousand instructions on average) than that of the service workloads (about 60 L2 cache misses per thousand instructions on average)  while higher than that of the HPCC workloads.
%The average L2 cache miss is lower than the all service workloads including media streaming,
%data severing, web search, web serving and SPEC web.
The L2 cache statistic indicates the data analysis workloads own better locality than the service workloads.
The HPCC workloads have different localities as the official web site mentioned, which can explain
the different cache behaviors among the HPCC workloads. %On the average, data analysis workloads has higher L2 cache misses that of the HPCC workloads.

From Figure \ref{l3cachel2}, we can find that for both the data analysis workloads and service workloads, the average ratio of L2 cache misses that are hit in L3 cache (85.5\% for the data analysis workloads and 94.9\% for the service workloads) is higher than that of the HPCC workloads. We can conclude that for most of the data analysis and service workloads, modern processor's LLC is large enough to cache most of data missed from L2 cache.

%Some of HPCC workloads whose temporal locality and spatial locality are very good, e.g. HPL,
%have less cache miss.
%For some HPCC workloads, whose locality is poor in either temporal or spatial, their have much cache
%miss, but for the data they processed are small when compared with big data analysis workloads, their cache miss can not exceed the big data applications.
%For HPC benchmarks, they all have few cache misses for most of the data
%, which they calculating, fit the
%cache size.

%We find the L2 cache misses may have impacts on the RS and  stall.
%The workloads, who owns a large amount of L2 cache misses allows have large percentage
%RAT stall in Figure \ref{stall}.  The RAT stall in pipeline is caused by
%the current instruction using a register that was partially written by previous instruction.
%While the L2 cache miss cause the long latency memory visits,
%which cause the allocated register waiting for the data coming.

\begin{figure}
\centering
\includegraphics[scale=0.45]{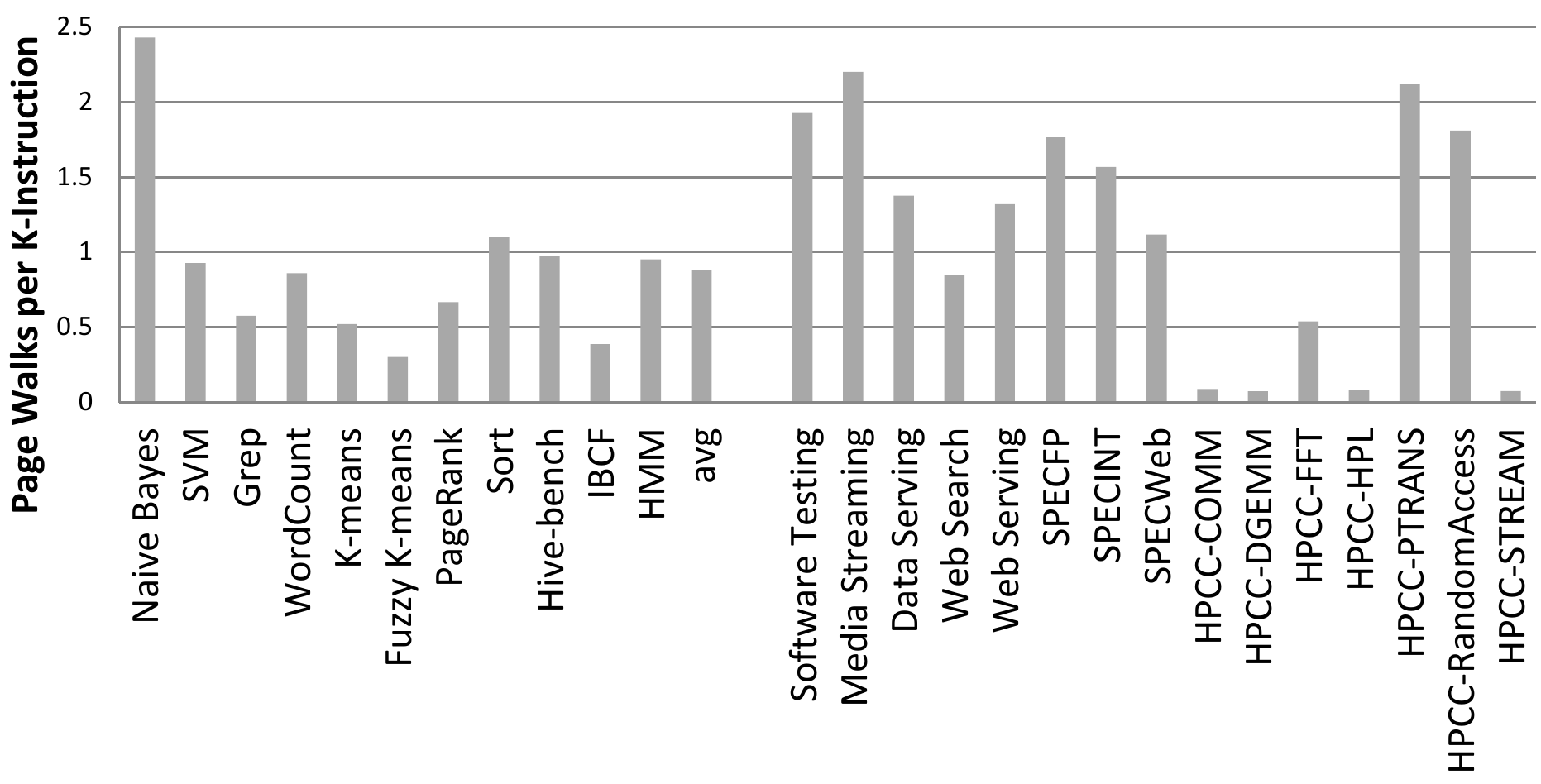}
\caption{Completed Page Walks Caused by DTLB Misses per Thousand Instructions Retired.}\label{dtlb}
\end{figure}

Figure \ref{dtlb} shows the completed page walks caused by data TLB misses per thousand instructions.
For most of the data analysis workloads with the exception of \emph{Naive Bayes}, the completed page walks caused by data TLB misses are  less than most of the service workloads and SPEC CPU2006 workloads (\emph{SPECINT} and \emph{SPECFP}), but higher than most of the HPCC workloads with the exception of \emph{HPCC-RandomAcess} and \emph{HPCC-PTRANS}.
That means the data locality of most of the data analysis workloads is much better than that of the service workloads.
%Most of data TLB entries do not need frequently changed, so the processor does not need many page walks.

%This observation is different from  that of instruction TLB. As mentioned in Section \ref{front}, the data analysis applications' completed page walks caused by instruction TLB miss are more frequently than that of SPECINT, SPECFP, and all HPCC workloads.
%Some service workloads (media streaming and data serving workloads) have more
%page walk than that of the data analysis workloads.

%other benchmark workloads
%and even less than the famous CPU benchmark workloads --SPECINT and SPECFP.
%The data analysis applications' ITLB behaviors and DTLB
%behaviors seem not balance, They incur more ITLB caused page walks than the traditional benchmark and less ITLB caused completed page walks than some service workloads. While they have less completed page walks caused by DTLB than nearly both traditional workloads and service workloads.

\emph{\textbf{Implications:}}

For the data analysis workloads, L2 cache is acceptably effective, and they have lower
L2 cache misses than that of the service workloads, while higher than that of the HPCC workloads.
%It only	has more L2 cache misses than HPCC workloads.
Meanwhile, for the data analysis and service workloads, most of L2 cache misses are hit in L3 cache, indicating L3 cache is pretty effective.
%iciency, decreasing the L3 cache size may have less impact on the whole performance.
%Considering that modern processors dedicate approximately half of the die
Modern processors dedicate approximately half of the die
area for caches, and hence optimizing the LLC capacity properly will improve the energy-efficiency of processor and save the die area.
For the service workloads, our observation corroborate the previous work \cite{ferdman2011clearing}: the L2 cache is ineffective.
\subsection{Branch Prediction}

The branch instruction prediction accuracy
is one of the most important factor that directly affects the
performance.
Modern out-of-order processors
introduce a functional unit (e.g. Branch Target Buffer)
to predict the next branch to avoid pipeline stalls due to branches.
If the predict is correct, the
pipeline will continue. However,
%if guess is wrong, it will cause a great latency.
if a branch instruction is mispredicted,
the pipeline must flush the wrong instructions and fetch the correct ones, which will cause at least a dozen of cycles' penalty.
So  branch prediction is not a trivial issue in the pipeline.

\begin{figure}
\centering
\includegraphics[scale=0.4]{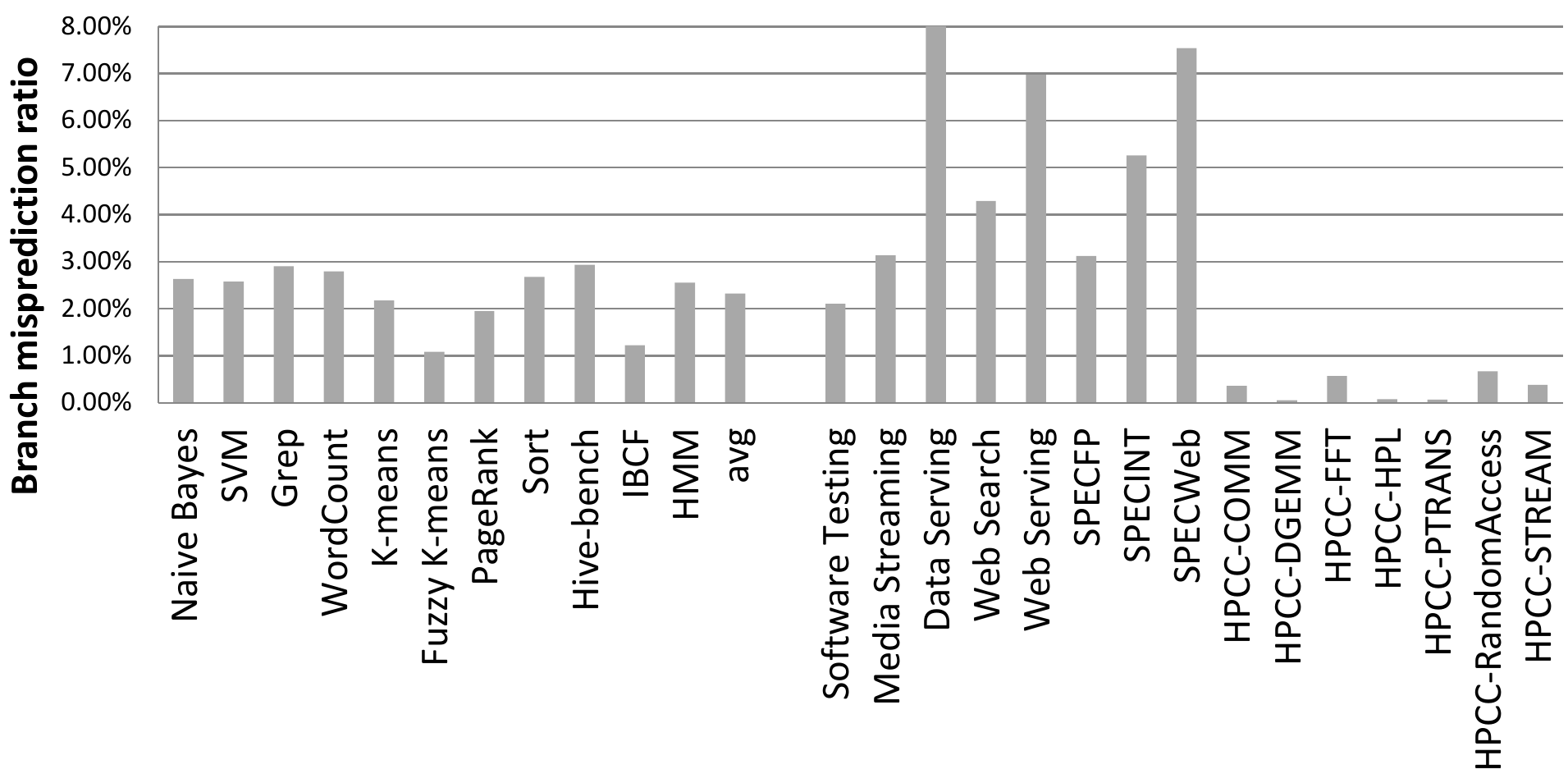}
\caption{Branch Miss-prediction ratio.}\label{branch-miss}
\end{figure}
Figure \ref{branch-miss} presents the branch miss prediction ratios of each workload.
We find that most of the data analysis workloads own a lower branch misprediction ratio in comparison with that of the service workloads and SPEC CPU workloads.
The HPCC workloads own very low misprediction ratios because the branch
logic codes of the seven micro
benchmarks are simple and the branch behaviors have great regularity.
%For instance the HPL (solving linear equations)
%only a few of workloads own high misprediction ratio.
The low misprediction ratios of the data analysis workloads indicate that most of the branch
instructions in the data analysis workloads have simple patterns. %which are easy to predict.
The simple patterns are conducive to BTB (Branch Target Buffer) to predict whether the next
branch needs to jump or not.
For  the data analysis workloads,  simple algorithms
chosen for big data always beat better
sophisticated algorithms\cite{rajaraman2008more}, which may be the possible  reason for their  low misprediction ratios.

\emph{\textbf{Implications: }}

Modern processors invest heavily in silicon real estate and algorithms for the branch
prediction unit in order to minimize the frequency and impact of wrong branch prediction.
For the data analysis workload, the misprediction ratio is lower than most of the compared workloads,
even for the CPU benchmark --- SPECINT, which implies that the branch predictor of modern processor
is good enough for the data analysis workloads. A simpler branch predictor may be preferred so
as to save power and die area.
\section{ The summary of DCBench}\label{dcbench}

Researchers in both academia and industry pay great attention to innovative date
center computer  systems and architecture.
Since benchmarks, as the foundation of quantitative design approach, are used
ubiquitously to evaluate the benefits of new designs and new
systems~\cite{bienia2011benchmarking},  we released a benchmark suite named \emph{DCBench} for datacenter computing  with an open-source license on our project home page on \url{http://prof.ict.ac.cn/DCBench}.

According to our workload characterization work, the  data analysis applications
share many inherent characteristics, different from traditional server
(SPECweb2005) and scale-out service workloads (four among
six benchmarks in CloudSuite), so DCBench includes two different kinds of workloads:  data analysis and service workloads.  In our research work, we
also notice the significant effects of different programming models, e. g., MPI vs. MapReduce,  on the application behaviors, which is beyond the scope of this paper, so we also include the implementation of DCBench with different programming models on our homepage. Other important factors include OS \cite{chenzheng} and VM executions, so we also
provide some VM images on our homepage for downloading.  We hope that \emph{DCBench} is helpful for performing  architecture and small-to-medium scale system researches for datacenter computing.

In addition to DCBench, we also release a big data benchmark suite---BigDataBench from Internet services \cite{gaoisca}, and a cloud computing benchmark suite---CloudRank \cite{luo2012cloudrank}. The purpose of BigDataBench is for large-scale system and architecture researches. We believe that the focus of cloud computing is to consolidate different workloads on a datacenter, which provides elastic resource management. So the goal of CloudRank is to model complex usage scenarios of cloud computing for the purpose of capacity planning, system evaluation and researches.

\section{Conclusion and Future Work}\label{conclusion}
%4.Different workloads have significantly varying architectural characteristics in terms of misprediction ratio, cache misses, pipeline stalls etc.,

In this paper, after investigating  most important application domains in terms of page views and daily visitors,
 %\emph{search engine, social networks, and electronic commerce},
 we chosen eleven representative data analysis workloads and  characterized their  micro-architectural characteristics on the systems equipped with modern superscalar out-of-order processors by using hardware performance counters.

 %,
%in order to understand the impacts and implications of data analysis
%workloads
%behind creating such a benchmark suite and then present the proposed benchmark
%suite, \emph{HVCBench}.  Our benchmark suite is extracted 21
%representative applications selected from massive data center applications.

Our work
shows that the data analysis workloads are significantly diverse in terms of both speedup performance  and micro-architectural characteristics.
% (Section \ref{characterization_BDbenchmarks}).
%  , implied that these applications have different
%user observation characteristics. In our experiments, some applications' speed up at 8 nodes are greater than 8, this is
%because that we use the same data set for all of the experiments, which resulted in the data set(around 150GB ) maybe too large for
%one slaves to process.
 %Meanwhile our  experiments in Section \ref{characterization_BDbenchmarks} also shows that different data analysis applications have
%different .
In a word, only one application is not enough to represent various categories of data analysis  workloads.

Our study on
the workloads reveals
%two unique aspects of data analysis applications in data center.
%First,
that the data analysis applications
%may have very different architectural characteristics, according which we give some
%implications to system engineers and processor researchers.
%they
share many inherent characteristics,
which place them in a different class from desktop, HPC,  traditional server and scale-out service
workloads, and accordingly we give several recommendations for architecture and system optimizations.  Meanwhile, we also observe  that the scale-out service workloads (four among six benchmarks in CloudSuite) share many similarity in terms of micro-architectural characteristics  with that of the traditional server workload characterized by SPECweb 2005. We will investigate more workloads to confirm this observation.

\section*{Acknowledgment}
We are very grateful to anonymous reviewers. This work is supported by the Chinese 973 project (Grant No.2011CB302502), the Hi-Tech Research and Development (863) Program of China (Grant No. 2011AA01A203,
 2013AA01A213), the NSFC project (Grant No.60933003, 61202075), the BNSF project (Grant No.4133081) and the 242 project (Grant No.2012A95).



%\bibliographystyle{IEEEtran}
%\begin{thebibliography}{1}
%\end{thebibliography}

%\newcommand{\BIBdecl}{\bfseries\setlength{\itemsep}{1\baselineskip plus 0.1\baselineskip minus 0.1\baselineskip}
\begin{small}
\bibliographystyle{IEEEtran}
%\linespread{0.5}
%\bibliographystyle{latex8}
%\setlength{\itemsep}{0pt}
\bibliography{IEEEabrv,tex}

\begin{thebibliography}{10}
\providecommand{\url}[1]{#1}
\csname url@samestyle\endcsname
\providecommand{\newblock}{\relax}
\providecommand{\bibinfo}[2]{#2}
\providecommand{\BIBentrySTDinterwordspacing}{\spaceskip=0pt\relax}
\providecommand{\BIBentryALTinterwordstretchfactor}{4}
\providecommand{\BIBentryALTinterwordspacing}{\spaceskip=\fontdimen2\font plus
\BIBentryALTinterwordstretchfactor\fontdimen3\font minus
  \fontdimen4\font\relax}
\providecommand{\BIBforeignlanguage}[2]{{%
\expandafter\ifx\csname l@#1\endcsname\relax
\typeout{** WARNING: IEEEtran.bst: No hyphenation pattern has been}%
\typeout{** loaded for the language `#1'. Using the pattern for}%
\typeout{** the default language instead.}%
\else
\language=\csname l@#1\endcsname
\fi
#2}}
\providecommand{\BIBdecl}{\relax}
\BIBdecl

\bibitem{zhan2012high}
J.~Zhan, L.~Zhang, N.~Sun, L.~Wang, Z.~Jia, and C.~Luo, ``High volume
  throughput computing: Identifying and characterizing throughput oriented
  workloads in data centers,'' in \emph{Parallel and Distributed Processing
  Symposium Workshops \& PhD Forum (IPDPSW), 2012 IEEE 26th
  International}.\hskip 1em plus 0.5em minus 0.4em\relax IEEE, 2012, pp.
  1712--1721.

\bibitem{barroso2009datacenter}
L.~Barroso and U.~H{\"o}lzle, ``The datacenter as a computer: An introduction
  to the design of warehouse-scale machines,'' \emph{Synthesis Lectures on
  Computer Architecture}, vol.~4, no.~1, pp. 1--108, 2009.

\bibitem{InfrastructureAtFacebook}
A.~Thusoo, Z.~Shao, S.~Anthony, D.~Borthakur, N.~Jain, J.~Sen~Sarma, R.~Murthy,
  and H.~Liu, ``Data warehousing and analytics infrastructure at facebook,'' in
  \emph{Proceedings of the 2010 international conference on Management of
  data}.\hskip 1em plus 0.5em minus 0.4em\relax ACM, 2010, pp. 1013--1020.

\bibitem{HadoopUsageReport}
\url{http://wiki.apache.org/hadoop/PoweredBy}.

\bibitem{zhancost}
J.~Zhan, L.~Wang, X.~Li, W.~Shi, C.~Weng, W.~Zhang, and X.~Zang, ``Cost-aware
  cooperative resource provisioning for heterogeneous workloads in data
  centers,'' \emph{Computers , IEEE Transactions on}.

\bibitem{wang2012cloud}
L.~Wang, J.~Zhan, W.~Shi, and Y.~Liang, ``In cloud, can scientific communities
  benefit from the economies of scale?'' \emph{Parallel and Distributed
  Systems, IEEE Transactions on}, vol.~23, no.~2, pp. 296--303, 2012.

\bibitem{sang2012precise}
B.~Sang, J.~Zhan, G.~Lu, H.~Wang, D.~Xu, L.~Wang, Z.~Zhang, and Z.~Jia,
  ``Precise, scalable, and online request tracing for multitier services of
  black boxes,'' \emph{Parallel and Distributed Systems, IEEE Transactions on},
  vol.~23, no.~6, pp. 1159--1167, 2012.

\bibitem{narayanan2006minebench}
R.~Narayanan, B.~Ozisikyilmaz, J.~Zambreno, G.~Memik, and A.~Choudhary,
  ``Minebench: A benchmark suite for data mining workloads,'' in \emph{Workload
  Characterization, 2006 IEEE International Symposium on}.\hskip 1em plus 0.5em
  minus 0.4em\relax Ieee, 2006, pp. 182--188.

\bibitem{huang2010hibench}
S.~Huang, J.~Huang, J.~Dai, T.~Xie, and B.~Huang, ``The hibench benchmark
  suite: Characterization of the mapreduce-based data analysis,'' in \emph{Data
  Engineering Workshops (ICDEW), 2010 IEEE 26th International Conference
  on}.\hskip 1em plus 0.5em minus 0.4em\relax IEEE, 2010, pp. 41--51.

\bibitem{ferdman2011clearing}
M.~Ferdman, A.~Adileh, O.~Kocberber, S.~Volos, M.~Alisafaee, D.~Jevdjic,
  C.~Kaynak, A.~Popescu, A.~Ailamaki, and B.~Falsafi, ``Clearing the clouds: A
  study of emerging workloads on modern hardware,'' \emph{Architectural Support
  for Programming Languages and Operating Systems}, 2012.

\bibitem{Alexa}
\url{http://www.alexa.com/topsites/global;0}, February, 2013.

\bibitem{bienia2008parsec}
C.~Bienia, S.~Kumar, J.~Singh, and K.~Li, ``The parsec benchmark suite:
  Characterization and architectural implications,'' in \emph{Proceedings of
  the 17th international conference on Parallel architectures and compilation
  techniques}.\hskip 1em plus 0.5em minus 0.4em\relax ACM, 2008, pp. 72--81.

\bibitem{dean2008mapreduce}
J.~Dean and S.~Ghemawat, ``Mapreduce: Simplified data processing on large
  clusters,'' \emph{Communications of the ACM}, vol.~51, no.~1, pp. 107--113,
  2008.

\bibitem{page1999pagerank}
L.~Page, S.~Brin, R.~Motwani, and T.~Winograd, ``The pagerank citation ranking:
  bringing order to the web.'' 1999.

\bibitem{hive-bench}
\url{https://issues.apache.org/jira/browse/HIVE\-396}.

\bibitem{hive}
\url{http://hive.apache.org/}.

\bibitem{mahout}
\url{http://mahout.apache.org/}.

\bibitem{cloudsuite}
\url{http://parsa.epfl.ch/cloudsuite/cloudsuite.html}.

\bibitem{cooper2010benchmarking}
B.~Cooper, A.~Silberstein, E.~Tam, R.~Ramakrishnan, and R.~Sears,
  ``{Benchmarking cloud serving systems with YCSB},'' in \emph{Proceedings of
  the 1st ACM symposium on Cloud computing}.\hskip 1em plus 0.5em minus
  0.4em\relax ACM, 2010, pp. 143--154.

\bibitem{faban}
``Faban harness and benchmark framework,'' \url{http://java.net/project/faban.}

\bibitem{keeton1998performance}
K.~Keeton, D.~A. Patterson, Y.~Q. He, R.~C. Raphael, and W.~E. Baker,
  \emph{Performance characterization of a quad Pentium pro SMP using OLTP
  workloads}.\hskip 1em plus 0.5em minus 0.4em\relax IEEE Computer Society,
  1998, vol.~26, no.~3.

\bibitem{eyerman2006performance}
S.~Eyerman, L.~Eeckhout, T.~Karkhanis, and J.~E. Smith, ``A performance counter
  architecture for computing accurate cpi components,'' in \emph{ACM SIGOPS
  Operating Systems Review}, vol.~40, no.~5.\hskip 1em plus 0.5em minus
  0.4em\relax ACM, 2006, pp. 175--184.

\bibitem{levinthal18027cycle}
D.~Levinthal, ``{Cycle accounting analysis on Intel Core 2 processors},''
  \emph{assets. devx. com/goparallel/18027. pdf}.

\bibitem{perf}
``Performance counters for linux,''
  \url{https://perf.wiki.kernel.org/index.php/Main_Page}.

\bibitem{intelref}
\emph{Intel 64 and IA-32 Architectures Software Developers Manual},
  Intel\textregistered, 2011.

\bibitem{thomadakis2011architecture}
M.~E. Thomadakis, ``The architecture of the nehalem processor and nehalem-ep
  smp platforms,'' \emph{Resource}, vol.~3, p.~2, 2011.

\bibitem{karkhanis2004first}
T.~S. Karkhanis and J.~E. Smith, ``A first-order superscalar processor model,''
  in \emph{Computer Architecture, 2004. Proceedings. 31st Annual International
  Symposium on}.\hskip 1em plus 0.5em minus 0.4em\relax IEEE, 2004, pp.
  338--349.

\bibitem{rajaraman2008more}
A.~Rajaraman, ``More data usually beats better algorithms,'' \emph{Datawocky
  Blog}, 2008.

\bibitem{bienia2011benchmarking}
C.~Bienia, ``Benchmarking modern multiprocessors,'' Ph.D. dissertation,
  Princeton University, 2011.

\bibitem{chenzheng}
Z.~Chen, Z.~Jianfeng, J.~Zhen, and Z.~Lixin, ``Characterizing os behavior of
  scale-out data center workloads,'' \emph{The Seventh Annual Workshop on the
  Interaction amongst Virtualization, Operating Systems and Computer
  Architecture (WIVOSCA 2013)}, 2013.

\bibitem{gaoisca}
W.~Gao and etc., ``Bigdatabench: a big data benchmark suite from web search
  engines,'' in \emph{The Third Workshop on Architectures and Systems for Big
  Data(ASBD 2013) in conjunction with The 40th International Symposium on
  Computer Architecture, May 2013.}

\bibitem{luo2012cloudrank}
C.~Luo, J.~Zhan, Z.~Jia, L.~Wang, G.~Lu, L.~Zhang, C.~Xu, and N.~Sun,
  ``Cloudrank-d: benchmarking and ranking cloud computing systems for data
  processing applications,'' \emph{Frontiers of Computer Science}, vol.~6,
  no.~4, pp. 347--362, 2012.

\end{thebibliography}
\end{small}

\end{document}